\documentclass[reprint,twocolumn,aps,showpacs,prb,superscriptaddress]{revtex4-2}
\usepackage{amsmath,amssymb,graphics,epsfig,epstopdf,color,multirow,array,verbatim,ulem,braket,tabularx,graphicx,color}
\usepackage[colorlinks,linkcolor=blue,citecolor=blue,urlcolor=blue]{hyperref}
\setcitestyle{square,numbers,sort&compress}

\begin{document}

\title{Electron- and hole-doped superconductivity in two-dimensional electride ZrX$_2$ (X = Cl, Br, I)}
\author{Tian-Xin Liu}
\affiliation{Department of Physics, School of Science, Jiangsu University of Science and Technology, Zhenjiang 212100, China}
\author{Xin-Yang Li}
\affiliation{Department of Physics, School of Science, Jiangsu University of Science and Technology, Zhenjiang 212100, China}
\author{Zuo-Wei Yin}
\affiliation{Department of Physics, School of Science, Jiangsu University of Science and Technology, Zhenjiang 212100, China}
\author{Jing Jiang}
\affiliation{School of Physics and Key Laboratory of Quantum State Construction and Manipulation (Ministry of Education), Renmin University of China, Beijing 100872, China}
\author{Ben-Chao Gong}\email{bcgong@just.edu.cn}
\affiliation{Department of Physics, School of Science, Jiangsu University of Science and Technology, Zhenjiang 212100, China}
\author{Wei Fan}
\affiliation{Department of Physics, School of Science, Jiangsu University of Science and Technology, Zhenjiang 212100, China}
\author{Xiao-Le Qiu}\email{qiu.xiaole@163.com}
\affiliation{School of Physics and Electronic Information, Weifang University, Weifang 261061, China}
\author{Kai Liu}
\affiliation{School of Physics and Key Laboratory of Quantum State Construction and Manipulation (Ministry of Education), Renmin University of China, Beijing 100872, China}

\begin{abstract}
Two-dimensional (2D) superconducting systems have recently emerged as a frontier in condensed matter physics, driven by their unique physical properties and potential applications. Using first-principles calculations, we show that monolayer ZrX$_2$ (X = Cl, Br, I), which is a family of 2D semiconducting electrides, hosts phonon-mediated superconductivity under both electron and hole dopings. Electron doping in ZrX$_{2}$ can achieve high superconducting transition temperatures at very low concentrations, with ZrBr$_2$ reaching $T_c = 30.0$ K at 0.01 \textit{e}/f.u., followed by ZrCl$_2$ (22.5 K at 0.05 \textit{e}/f.u.) and ZrI$_2$ (18.1 K at 0.01 \textit{e}/f.u.), for which the high $T_c$ originates from the flat conduction band minimum, with Van Hove singularities nearby enhancing the density of states. In contrast, hole doping requires higher concentrations (0.2--0.25 \textit{h}/f.u.) to activate superconductivity, yielding $T_c \sim 2$ K, and approaches a magnetic instability at elevated doping levels. By comparing with InSe and LaBr$_2$, we identify the spatial location and orbital character of excess electrons as key factors determining the strength of hole-doped superconductivity and the nature of competing orders. These findings establish monolayer ZrX$_2$ as a platform for bipolar superconductivity in a 2D electride and provide insights into flat-band engineering for high-$T_c$ superconductors.
\end{abstract}

\maketitle

\section{INTRODUCTION}

A hallmark of cuprate and iron-based high-temperature superconductors is the dome-shaped superconducting region on both the electron- and hole-doped sides of the phase diagram \cite{cuprate1,cuprate2,cuprate3,cuprate4,iron1,iron2,iron3,iron4}. Understanding the origin of such bipolar superconducting domes remains a central challenge, motivating the search for simpler material platforms where analogous behavior can be studied.

Besides the cuprate and iron-based superconductors with layered structures, two-dimensional (2D) semiconductors have also attracted intense interest recently due to their ability to host exotic quantum phases and their potential for quantum device applications \cite{detector1,flex2}. In past decades, various strategies have been developed to induce superconductivity in these materials, including strain engineering \cite{mos2bts,asbts}, constructing van der Waals heterostructures (such as SnSe$_2$/graphene \cite{snse2graphene23} and SnSe$_2$/PtTe$_2$ \cite{snse2fan24}), intercalating or adatom adsorption (such as adsorbing alkali metal atoms on MoS$_2$ \cite{limos2}, ZrNCl \cite{lizrncl}, and InSe \cite{nainse}), as well as electron doping \cite{inseelectron30,phosphorene29}. Among them, carrier doping offers a particularly clean and tunable means to introduce charge carriers without altering the crystal structure or introducing disorder. However, the response to hole doping often differs markedly from that to electron doping. In many 2D semiconductors, hole doping tends to induce ferromagnetism rather than superconductivity \cite{magadd1,magadd2,magadd3,zrcl2mag34}. Consequently, the observation of superconductivity on both carrier types in 2D semiconductors is exceedingly rare.

Among various 2D semiconductors, monolayer InSe is a striking exception: electron doping yields a superconducting $T_c$ of 3.4 K at 0.2 \textit{e}/cell \cite{inseelectron30}, while hole doping achieves a remarkably high $T_c$ of 55--75 K \cite{insehole31}. What makes InSe so special? From an oxidation-state perspective, each unit contains two In and two Se atoms, leaving two excess electrons, which suggests its electron-rich properties. Moreover, the electrons in the In-In $\sigma$ bonds (Fig. \ref{fig.1}(a)), form a flat band at the valence band maximum (VBM) (Fig. \ref{fig.1}(c)). Hole doping directly metallizes this flat band, producing a high density of states at the Fermi level and strong electron-phonon coupling \cite{sigmabond,cnphysics}. This observation raises the compelling question: can other electron-rich 2D materials exhibit similar behavior, and what determines whether hole doping yields high-$T_c$ superconductivity or magnetism?

Electrides---materials where excess electrons occupy interstitial sites rather than atomic orbitals \cite{van_ene3,ca2ncal6,ba2ncal13}---offer an ideal testing ground. Monolayer LaBr$_2$, a 2D electride with one excess electron per formula unit, exhibits intrinsic ferromagnetism due to half-filling of an interstitial flat band \cite{labr2prb35,labr2val36}. For electrides with two excess electrons per formula unit, such as the ZrX$_2$ (X = Cl, Br, I) series \cite{zrcl2mag34,zrx2cal61,2zrx2}, the interstitial band is fully occupied, yielding nonmagnetic (NM) semiconductors. Previous studies have investigated the valley polarization \cite{2zrx2}, thermoelectric \cite{thelezri239}, and transport properties \cite{nanozri240} of ZrX$_2$ monolayers, but their superconducting properties remain unexplored.

In this work, we address these questions through systematic first-principles calculations on monolayer ZrX$_2$ (X = Cl, Br, I). Our results reveal that these 2D electrides exhibit superconductivity on both electron- and hole-doped sides. Electron doping achieves high $T_c$ at very low concentrations, mirroring the mechanism of hole-doped InSe: the conduction band minimum of ZrX$_2$ is intrinsically flat. Hole doping, in contrast, requires higher concentrations to activate superconductivity and approaches a magnetic instability at elevated doping levels. By comparing with InSe and LaBr$_2$, we show that the spatial location of excess electrons determines the energy position of flat bands relative to band edges, which in turn governs the strength of superconductivity and the nature of competing orders.

\section{COMPUTATIONAL METHODS}

Density functional theory (DFT) calculations were performed by using the Quantum ESPRESSO package \cite{qe44,qe45,qe46}. The projected augmented wave (PAW) pseudopotentials \cite{paw41,paw42} and the Perdew-Burke-Ernzerhof (PBE) exchange-correlation functional \cite{gga43} were used. The kinetic energy cutoff for plane wave basis was set to 70 Ry, and the charge density cutoff to 400 Ry. The Fermi surface broadening was treated with Gaussian smearing of width 0.0037 Ry (0.05 eV). A vacuum layer exceeding 20 \AA \ was introduced to eliminate interactions between periodic images. Structural relaxations were converged to forces below $1 \times 10^{-8}$ Ry/bohr, and a $32 \times 32 \times 1$ Monkhorst-Pack $\mathbf{k}$-point mesh was used for Brillouin zone sampling.

Phonon dispersions and electron-phonon coupling (EPC) were calculated using density functional perturbation theory (DFPT) \cite{dfpt47,dfpt48} within the linear response regime. A $8 \times 8 \times 1$ $\mathbf{q}$-point mesh was employed, along with a sparse $16 \times 16 \times 1$ $\mathbf{k}$-point mesh and two dense $\mathbf{k}$-point meshes: $64 \times 64 \times 1$ and $96 \times 96 \times 1$. The superconducting transition temperature $T_c$ was evaluated using the McMillan-Allen-Dynes formula \cite{allen49}:
\begin{equation}
	T_\text{c} = \dfrac{\omega_{\text{log}}}{1.2} \exp\left[\dfrac{-1.04(1+\lambda)}{\lambda (1-0.62 \mu^{*})-\mu^{*}}\right],
\label{eq:mcMillan}
\end{equation}
where $\omega_{\text{log}}$ is the logarithmic average phonon frequency, $\mu^{*} = 0.1$ is the effective screened Coulomb repulsion constant \cite{miu50,miu51}, and $\lambda$ is the total EPC constant which can be calculated by integrating the Eliashberg spectral function $\alpha^{2} F(\omega)$ or summing the EPC constants $\lambda_{\textbf{q} \nu}$. 
The Eliashberg spectral function $\alpha^2F(\omega)$ and phonon linewidth $\gamma_{\textbf{q}\nu}$ are defined as:
\begin{equation}
	\lambda = 
	2 \int_{}^{} \dfrac{\alpha^{2} F(\omega)}{\omega} d \omega = \sum_{\textbf{q} \nu}^{} \lambda_{\textbf{q} \nu},
	\label{eq:lambda}
\end{equation}
\begin{equation}
	\alpha^{2} F(\omega) = \dfrac{1}{2 \pi N(\varepsilon_{F})} \sum_{\textbf{q} \nu} \delta (\omega - \omega_{\textbf{q} \nu}) \dfrac{\gamma_{\textbf{q} \nu}}{\omega_{\textbf{q} \nu} \hbar},
	\label{eq:alpha2F}
\end{equation}
\begin{equation}
	\gamma_{\textbf{q} \nu} = 2 \pi \omega_{\textbf{q} \nu} \sum_{\textbf{k}nn^{\prime}} \left| g^{\textbf{q} \nu}_{\textbf{k}+\textbf{q}n^{\prime},\textbf{k}n} \right|^{2} \delta (\varepsilon_{\textbf{k}n}-\varepsilon_{F}) \delta(\varepsilon_{\textbf{k}+\textbf{q}n^{\prime}}-\varepsilon_{F}),
	\label{eq:gamma}
\end{equation}
where $N(\varepsilon_{F})$ is the density of states at the Fermi level, $\omega_{\textbf{q} \nu}$ is the frequency of the $\nu$th phonon mode at wavevector $\mathbf{q}$, and $g^{\textbf{q} \nu}_{\textbf{k}+\textbf{q}n^{\prime},\textbf{k}n}$ denotes the EPC matrix element.

The \textit{ab initio} molecular dynamics (AIMD) simulations were carried out at 300 K using canonical ensemble (NVT) with a time step of 2 fs. 
The convergence tests for the kinetic-energy cutoff, charge-density cutoff, SCF Fermi-surface smearing width (degauss), electron-phonon Gaussian broadening parameter $\sigma_{el-ph}$ (el\_ph\_sigma) and q-mesh are provided in the Supplemental Material (Figs. S1 and S2) \cite{supplement}, where the DFPT results are also verified against EPW calculations \cite{epw1,epw2,epw3}.

\section{RESULTS AND DISCUSSION}

\subsection{Crystal structure and stability}

\begin{figure}[htbp]
	\centering
	\includegraphics[width=0.97\columnwidth]{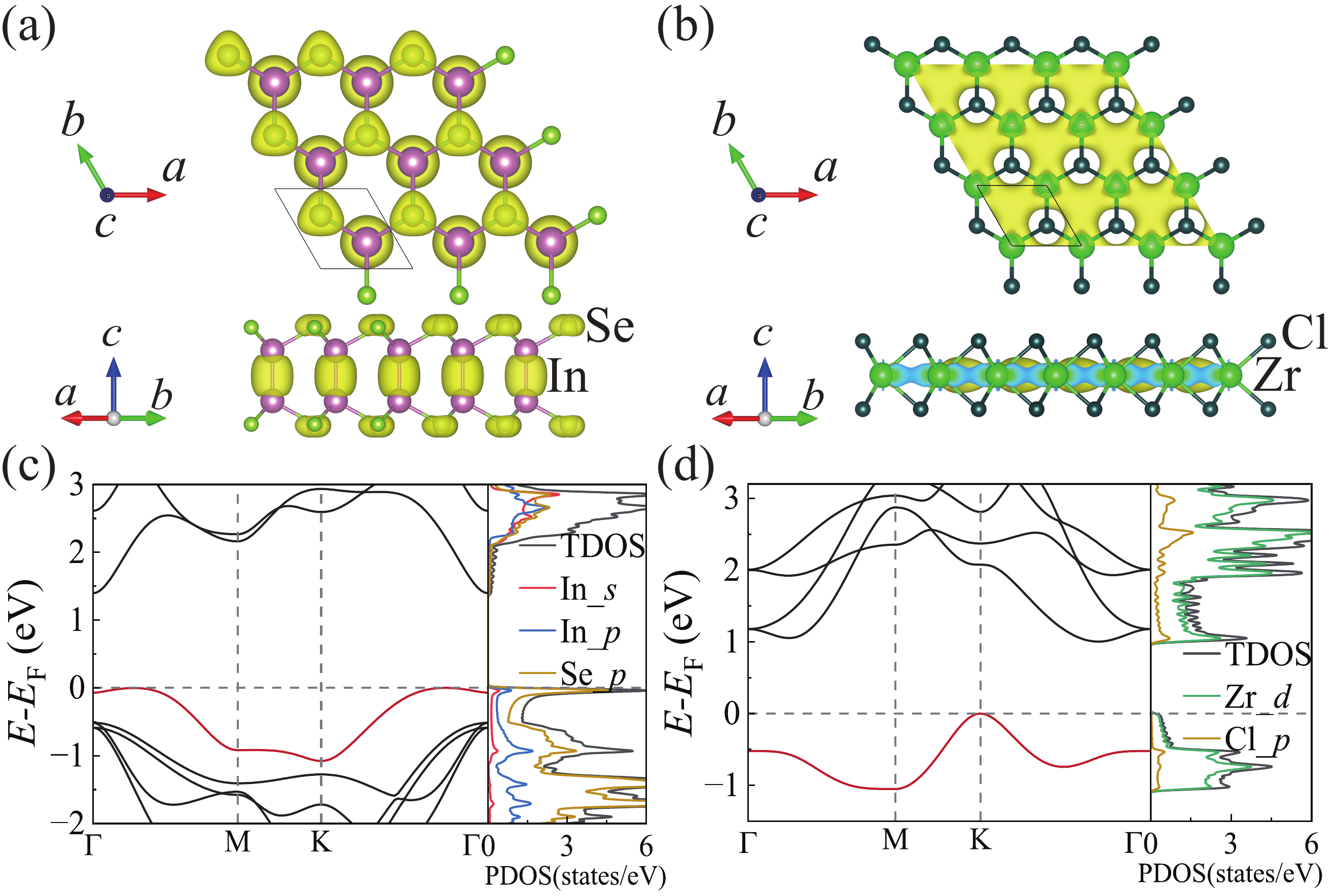}
	\caption{Top and side views of monolayer (a) InSe and (b) ZrCl$_2$. The yellow isosurface indicates the integrated charge densities (ICDs) of the topmost valence band. The isosurface of charge densities is set to 0.008 e/Å$^{3}$. Pink, light green, green, and blue-green balls represent In, Se, Zr, and Cl atoms, respectively. Band structure and density of states (DOS) of (c) InSe and (d) ZrCl$_2$. The topmost valence band is highlighted in red.}
	\label{fig.1}
\end{figure}

Monolayer ZrX$_2$ (X = Cl, Br, I) crystallizes in the H-phase structure with space group $P\bar{6}m2$ and point group $D_{3h}$, as illustrated in Fig. \ref{fig.1}(b) for ZrCl$_2$. The structure consists of X-Zr-X trilayers stacked in a hexagonal lattice, analogous to H-phase MoS$_2$ \cite{mos2}. Each Zr atom is coordinated by six X atoms in a prismatic geometry, which splits the Zr $d$ orbitals into three groups: $a$ ($d_{z^2}$), $e_1$ ($d_{xz}$, $d_{yz}$), and $e_2$ ($d_{xy}$, $d_{x^2-y^2}$), consistent with previous reports \cite{zrcl2mag34}. The optimized lattice constants are $a = 3.41$ \AA \ for ZrCl$_2$, $a = 3.56$ \AA \ for ZrBr$_2$, and $a = 3.83$ \AA \ for ZrI$_2$, in good agreement with previous theoretical reports \cite{zrx2cal61,2zrx2}. All subsequent doping calculations are performed using these optimized lattice constants, while only the atomic coordinates are relaxed.

Bulk 3R-ZrX$_2$ (space group $R_{3m}$, see Supplemental Material Fig. S3(a)) \cite{supplement} has been known since the 1970s \cite{zrclexp60}. The cleavage energy, calculated by separating the topmost layer from a six-layer slab, is 0.300 J/m$^2$ for ZrCl$_2$, 0.318 J/m$^2$ for ZrBr$_2$, and 0.317 J/m$^2$ for ZrI$_2$ (Fig. S3(b)) \cite{supplement}. These values are significantly lower than that of graphene (0.36 J/m$^2$) \cite{cleagra64,cleagra65}, suggesting that monolayer ZrX$_2$ can be readily exfoliated. The stability of the monolayer is confirmed by phonon calculations. The absence of imaginary frequencies in the phonon dispersions (Figs. S3(c), S4(b) and S4(d)) \cite{supplement} confirms the dynamical stability. Meanwhile, the AIMD simulations show that monolayers ZrX$_2$ (X = Cl, Br, I) are thermodynamically stable at room temperature (Fig. S5) \cite{supplement}. 

\subsection{Electronic structure and flat bands}

From a simple valence-electron count, each Zr atom has four valence electrons and the two halogen atoms require two electrons in total, leaving two excess electrons per formula unit. These excess electrons fill the highest valence band, opening a band gap of 1.00\,eV (ZrCl$_2$), 0.85\,eV (ZrBr$_2$), and 0.76\,eV (ZrI$_2$) (Figs.~\ref{fig.1}(d), S4(a) and S4(c)) \cite{supplement}. The integrated charge densities (ICDs) of topmost valance band (Fig.~\ref{fig.1}(b)) shows that these electrons localize at hexagonal interstitial sites, a hallmark of two-dimensional electrides~\cite{van_ene3}. The electride character is further evidenced by the electron localization function (ELF) in Fig. S12 and Table SI \cite{supplement}. Hence, monolayer ZrX$_2$ can be formulated as $[\text{ZrX}_2]^{2+} \cdot 2e^-$~\cite{zrcl2mag34}.

In the band structure, the VBM at K derives mainly from the degenerate in-plane $d_{xy}/d_{x^2-y^2}$ orbitals (Fig. S6) \cite{supplement}. At $\Gamma$, a band with weaker dispersion and a higher DOS appears $\sim$0.5\,eV below the VBM (Fig.~\ref{fig.1}(d)), originating from the out-of-plane $d_{z^2}$ orbital (Fig. S6) \cite{supplement}. Near the conduction-band minimum (CBM), the band exhibits six equivalent Mexican-hat-like features around $\Gamma$ point, forming a weakly dispersive ring near the CBM (Fig. S7). Then, it can be regarded as a flat band and produces a sharp peak in the density of states (Fig.~\ref{fig.1}(d)). These weakly dispersive bands with large DOS may strengthen electron--phonon coupling and promote electronic instabilities, upon carrier doping~\cite{insehole31,zrcl2mag34}.

\subsection{Electron-doped superconductivity}

\begin{figure}[htbp]
	\centering
	\includegraphics[width=0.97\columnwidth]{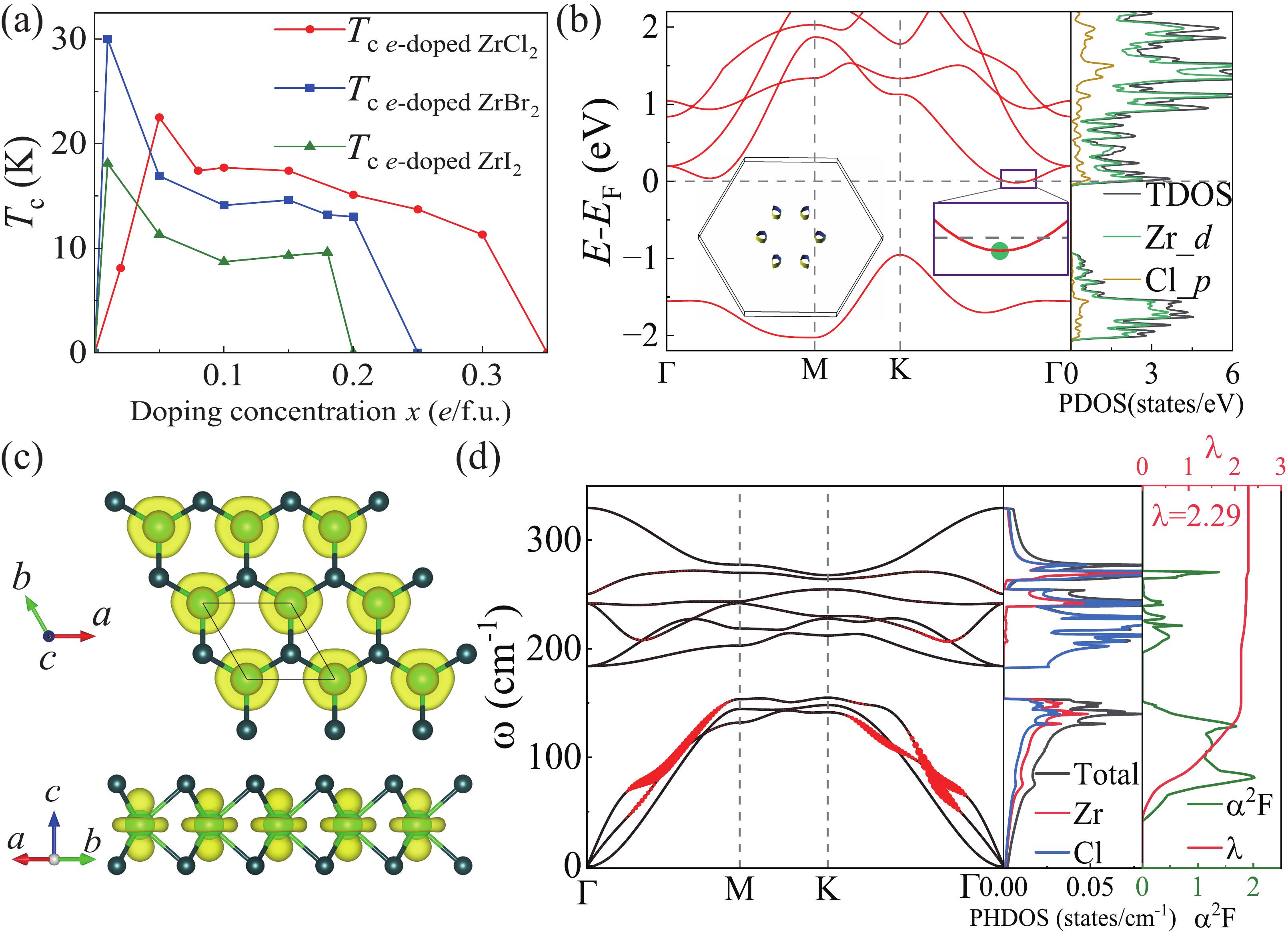}
	\caption{(a) Superconducting $T_c$ of monolayer ZrX$_2$ (X = Cl, Br, I) as functions of electron doping concentration $x$. (b) Band structure, density of states and Fermi surface of monolayer ZrCl$_2$ at $x = 0.05$ \textit{e}/f.u. (c) The ICDs of the CBM, corresponding to the region indicated by green dots in the enlarged view of (b). The isosurface of charge densities is set to 0.015 e/Å$^{3}$. (d) Phonon dispersion, phonon density of states, Eliashberg spectral function $\alpha^2F(\omega)$, and electron-phonon coupling constant $\lambda$ of monolayer ZrCl$_2$ at $x = 0.05$ \textit{e}/f.u.}
	\label{fig.2}
\end{figure}

We first study electron-doped ZrX$_2$ by changing the total number of electrons with a compensating jellium background. Figure \ref{fig.2}(a) displays the superconducting $T_c$ as a function of electron doping concentration $x$ for all ZrX$_2$ compounds. Superconductivity emerges at remarkably low doping levels: $T_c$ becomes nonzero at $x = 0.01$ \textit{e}/f.u. for ZrBr$_2$ and ZrI$_2$, and at $x = 0.02$ \textit{e}/f.u. for ZrCl$_2$. For all three compounds, $T_c$ exhibits a dome-shaped dependence on doping. ZrBr$_2$ achieves the highest $T_c$ of 30.0 K at $x = 0.01$ \textit{e}/f.u., followed by ZrCl$_2$ (22.5 K at $x = 0.05$ \textit{e}/f.u.) and ZrI$_2$ (18.1 K at $x = 0.01$ \textit{e}/f.u.). Beyond the optimal doping, $T_c$ decreases monotonically. At sufficiently high doping ($x = 0.35$ \textit{e}/f.u. for ZrCl$_2$), imaginary frequencies appear in the phonon dispersion (Fig. S3(d)) \cite{supplement}, indicating lattice instability.

Unlike electron-doped InSe, which exhibits a relatively low $T_c$ of 3.4 K \cite{inseelectron30} (Table \ref{tab.1}), electron-doped ZrX$_2$ compounds achieve much higher superconducting transition temperatures (up to 30.0 K in ZrBr$_2$). The origin of this enhancement can be traced to the electronic structure near the Fermi level. Taking ZrCl$_2$ at its optimal doping ($x = 0.05$~\textit{e}/f.u.) as a representative case, Fig.~\ref{fig.2}(b) displays the band structure and projected density of states (PDOS). The Fermi level shifts into the conduction band, forming six electron pockets centered at $\Gamma$ (inset of Fig.~\ref{fig.2}(b)). The CBM of pristine ZrCl$_2$, dominated by Zr $d_{z^2}$ orbitals (Fig.~\ref{fig.2}(c)), is surrounded by a nearly flat ring arising from six Van Hove singularities (Fig. S7), leading to a sharp DOS peak at $E_\text{F}$ with $N(\varepsilon_\text{F}) = 18.10$~states/eV/spin/f.u.. This value is substantially larger than that of electron-doped InSe ($N(\varepsilon_\text{F}) = 7.70$~states/eV/spin/f.u.)~\cite{inseelectron30}.

The electron-phonon coupling at this doping level is presented in Fig. \ref{fig.2}(d). The absence of imaginary frequencies confirms dynamic stability. The EPC strength, reflected in the phonon linewidths $\gamma_{\textbf{q}\nu}$, is dominated by acoustic phonon branches in the 50--150 cm$^{-1}$ range. Their integration yield $\lambda = 2.29$, with over 90\% of this value originating from low-frequency acoustic phonons, also larger than that of electron-doped InSe ($\lambda = 0.65$) \cite{inseelectron30}. Similar behavior is observed for ZrBr$_2$ and ZrI$_2$ (Fig. S8) \cite{supplement}. Thus, the combination of high density of states at the Fermi level and strong coupling to low-frequency acoustic phonons accounts for the high $T_c$ in electron-doped ZrX$_2$. Considering the large $\lambda$ in electron-doped ZrX$_2$, we also calculated $T_c$ with the Allen–Dynes modified formula \cite{ad}. The corrected values are slightly higher but show the same doping dependence (see Fig. S9 in the Supplemental Material). 

\subsection{Hole-doped superconductivity}
\begin{figure}[htbp]
	\centering
	\includegraphics[width=0.97\columnwidth]{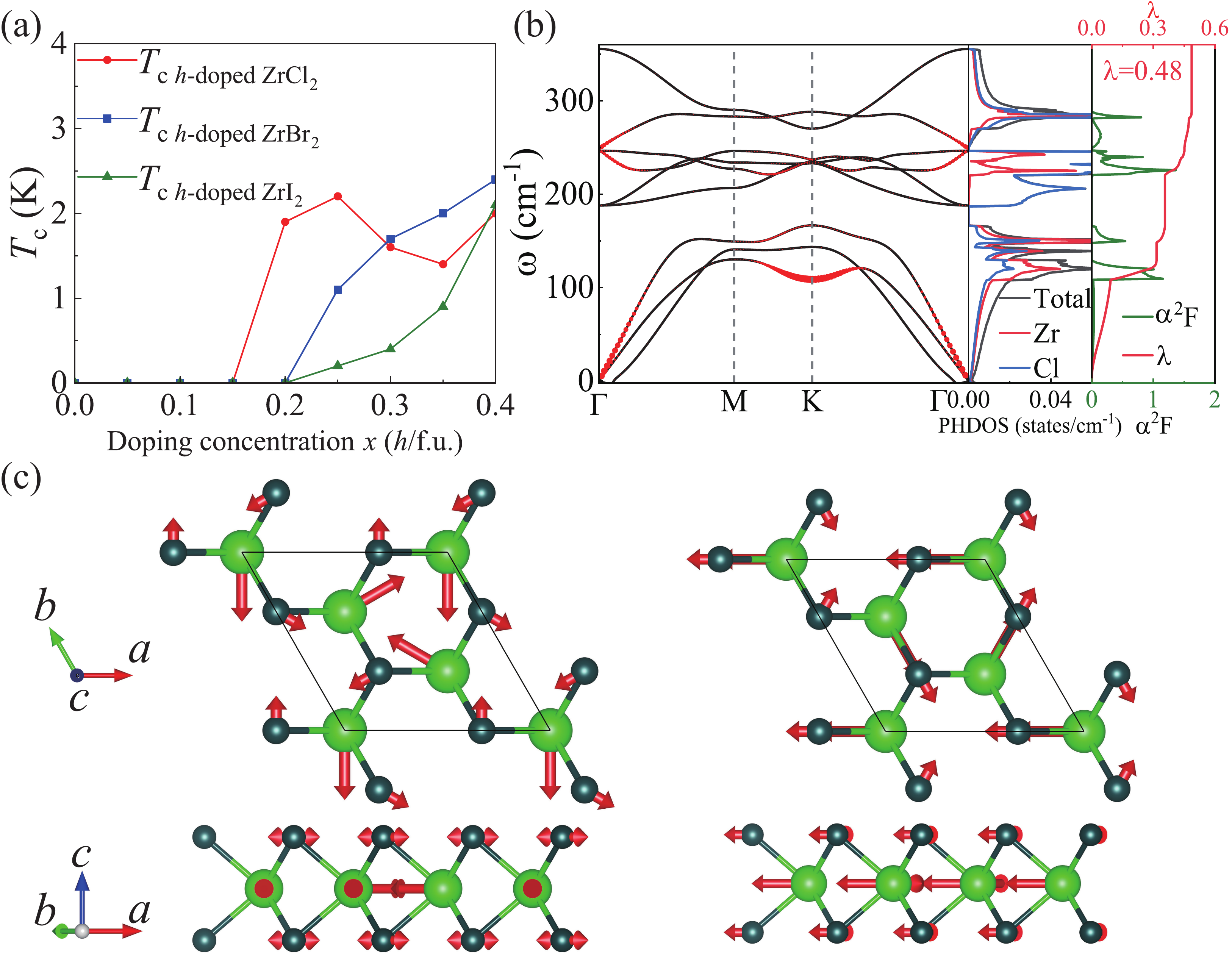}
	\caption{(a) Superconducting $T_c$ of monolayer ZrX$_2$ (X = Cl, Br, I) as functions of hole doping concentration $x$. (b) Phonon dispersion and $\alpha^2F(\omega)$ of ZrCl$_2$ at $x = 0.2$ \textit{h}/f.u. The size of red dot is proportional to the phonon linewidth $\gamma_{\textbf{q}\nu}$. (c) Doubly degenerate phonon mode at K point ($\omega$ = 115 cm$^{-1}$), obtained from a $\sqrt{3}\times\sqrt{3}\times1$ supercell.}
	\label{fig.3}
\end{figure}

\begin{figure}[htbp]
	\centering
	\includegraphics[width=0.97\columnwidth]{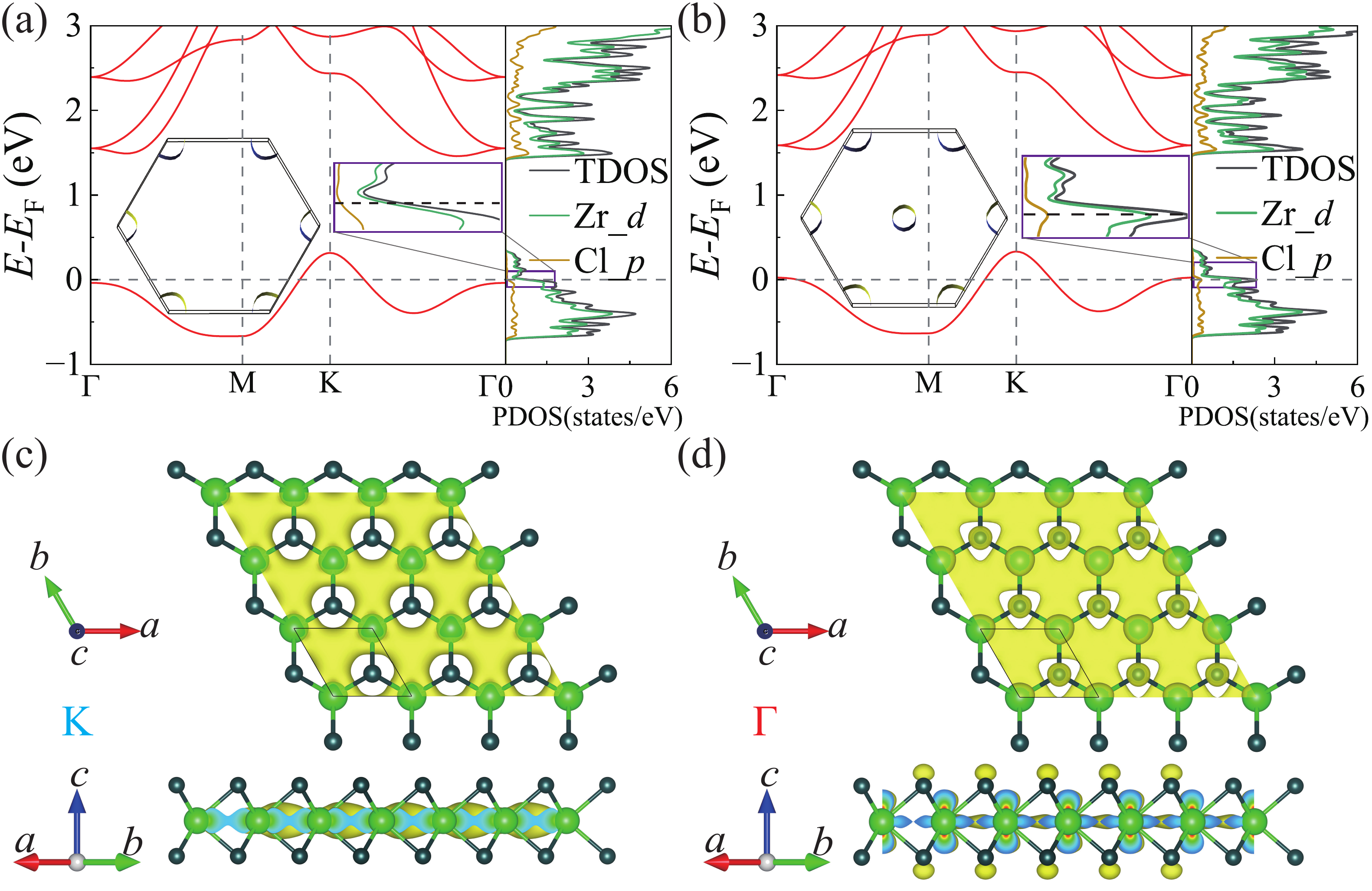}
	\caption{Band structure and Fermi surface of ZrCl$_2$ at (a) $x = 0.15$ \textit{h}/f.u. and (b) $x = 0.2$ \textit{h}/f.u. The ICDs of topmost valance band from (c) K point and (d) $\Gamma$ point. The isosurface of charge densities is set to 0.01 e/Å$^{3}$.}
	\label{fig.4}
\end{figure}

Hole doping leads to markedly different behavior. Figure \ref{fig.3}(a) shows superconducting $T_c$ of ZrX$_2$ as functions of hole doping concentration $x$.  Obviously, the $T_c$ values are significantly lower than those achieved by electron doping. To understand this, we calculated the electron-phonon coupling of ZrCl$_2$ at $x = 0.2$ \textit{h}/f.u., as shown in Fig. \ref{fig.3}(b). The EPC strength is dominated by acoustic phonon branches in the 105--125 cm$^{-1}$ range, localized near the K point. Using a $\sqrt{3}\times\sqrt{3}\times1$ supercell, these modes are identified as a pair of doubly degenerate phonon modes (Fig. \ref{fig.3}(c)), corresponding to in-plane vibrations of Zr and Cl atoms. The total EPC constant is $\lambda = 0.48$, an order of magnitude smaller than the electron-doped case ($\lambda = 2.29$). In addition, the DOS at the Fermi level $N(\varepsilon_\text{F})$ of 10.10 states/eV/spin/f.u. (Fig. \ref{fig.4}(b)) at this doping level is also much lower than that of the electron-doped case ($\sim 18.10$ states/eV/spin/f.u. at $x = 0.05$ \textit{e}/f.u.). Thus, these all give rise to the low $T_c$ in hole-doped ZrCl$_2$. 

\begin{table*}[htbp]
	\centering
	\caption{Comparison of electronic and magnetic properties of InSe, LaBr$_2$, and ZrX$_2$.}
	\begin{tabular}{lccc}
		\hline
		& InSe & LaBr$_2$ & ZrX$_2$ (in this work) \\
		\hline
		Excess electrons each unit cell & 2 & 1 & 2 \\
		Electron localization & In-In $\sigma$ bonds & Hexagonal interstitial & Hexagonal interstitial \\
		Flat band location & VBM (band edge) & -- & CBM (band edge) \\
		Ground state & NM insulator & FM insulator \cite{labr2prb35} & NM insulator \\
		Hole-doped $T_c$ (K) & 55--75 \cite{insehole31} & -- & $\sim$2 \\
		Electron-doped $T_c$ (K) & 3.4 \cite{inseelectron30} & -- & 18--30 \\
		$\lambda$ (optimal doping) & $\sim$7.3--7.6 \cite{insehole31} & -- & 2.29 (ZrCl$_2$), 4.12 (ZrBr$_2$), 2.23 (ZrI$_2$) \\
		Competing order & Charge density wave \cite{insehole31} & -- & Spin density wave \\
		\hline
	\end{tabular}
	\label{tab.1}
\end{table*}

Meanwhile, superconductivity does not appear at low doping levels, and a threshold concentration is required: $x = 0.2$ \textit{h}/f.u. for ZrCl$_2$, and $x = 0.25$ \textit{h}/f.u. for ZrBr$_2$ and ZrI$_2$ (Figs. \ref{fig.3}(a)).  To further clarify this threshold, we examine the evolution of band structure around the superconducting critical point. Figures \ref{fig.4}(a) and \ref{fig.4}(b) show the band structures of ZrCl$_2$ at $x = 0.15$ \textit{h}/f.u. (below threshold) and $x = 0.2$ \textit{h}/f.u. (at threshold). At $x = 0.15$ \textit{h}/f.u., the Fermi level lies above the $\Gamma$-point band, and only a hole pocket at K point is present. At $x = 0.2$ \textit{h}/f.u., the Fermi level reaches the $\Gamma$-point band, forming an additional pocket where the flat $d_{z^2}$ band resides. This leads to a sharp jump in the density of states at the Fermi level, as clearly seen in the DOS curves of Figs. \ref{fig.4}(a) and \ref{fig.4}(b).

The ICDs in Figs. \ref{fig.4}(c) and \ref{fig.4}(d) reveal the orbital character of these states. At the K point (Fig. \ref{fig.4}(c)), the ICDs show electrons localized in the Zr-Zr plane and the interstitial region, with the latter corresponding to unbonding states. At the $\Gamma$ point (Fig. \ref{fig.4}(d)), the ICDs show hybridization between the Zr $d_{z^2}$ and Cl $p$ orbitals, indicating bonding states. Similar evolution is observed for ZrBr$_2$ and ZrI$_2$ (Figs. S10 and S11) \cite{supplement}. Comparing with the orbital-projected bands in Figs. S13(a) and S13(c), it suggests that the interstitial states are derived mainly from the in-plane Zr $d_{xy}$/$d_{x^2-y^2}$ orbitals. Hence, the interstitial unbonding electrons contribute negligibly to superconductivity, similar to monolayer electride Ba$_2$N \cite{ba2ncal13}. Superconductivity appears only when the Fermi level reaches the $\Gamma$-point band, activating the Zr $d_{z^2}$ and Cl $p$ orbitals, which show large response to the phonon modes in Fig. \ref{fig.3}(c). To further confirm this, we calculated the doping dependence of the PDOS/TDOS ratio at \textit{E}$_\text{F}$ from the interstitial quasi-atom (labled Q) and the Zr $d_{z^2}$ orbital (Fig. S14) \cite{supplement}. With increasing hole doping, the interstitial contribution decreases rapidly, while the Zr $d_{z^2}$ contribution increases and exceeds the interstitial contribution in the superconducting regime. This directly suggests that it is the Zr-$d_{z^2}$ rather than weak bonding interstitial electrons contribute to superconductivity in the hole doped ZrCl$_2$. This is consistent with Li-B-C compounds, in which states with weak bonding character exhibit small deformation potentials and consequently low $T_c$ \cite{gu2026}.

\subsection{Proximity to magnetism}

\begin{figure}[htbp]
	\centering
	\includegraphics[width=0.97\columnwidth]{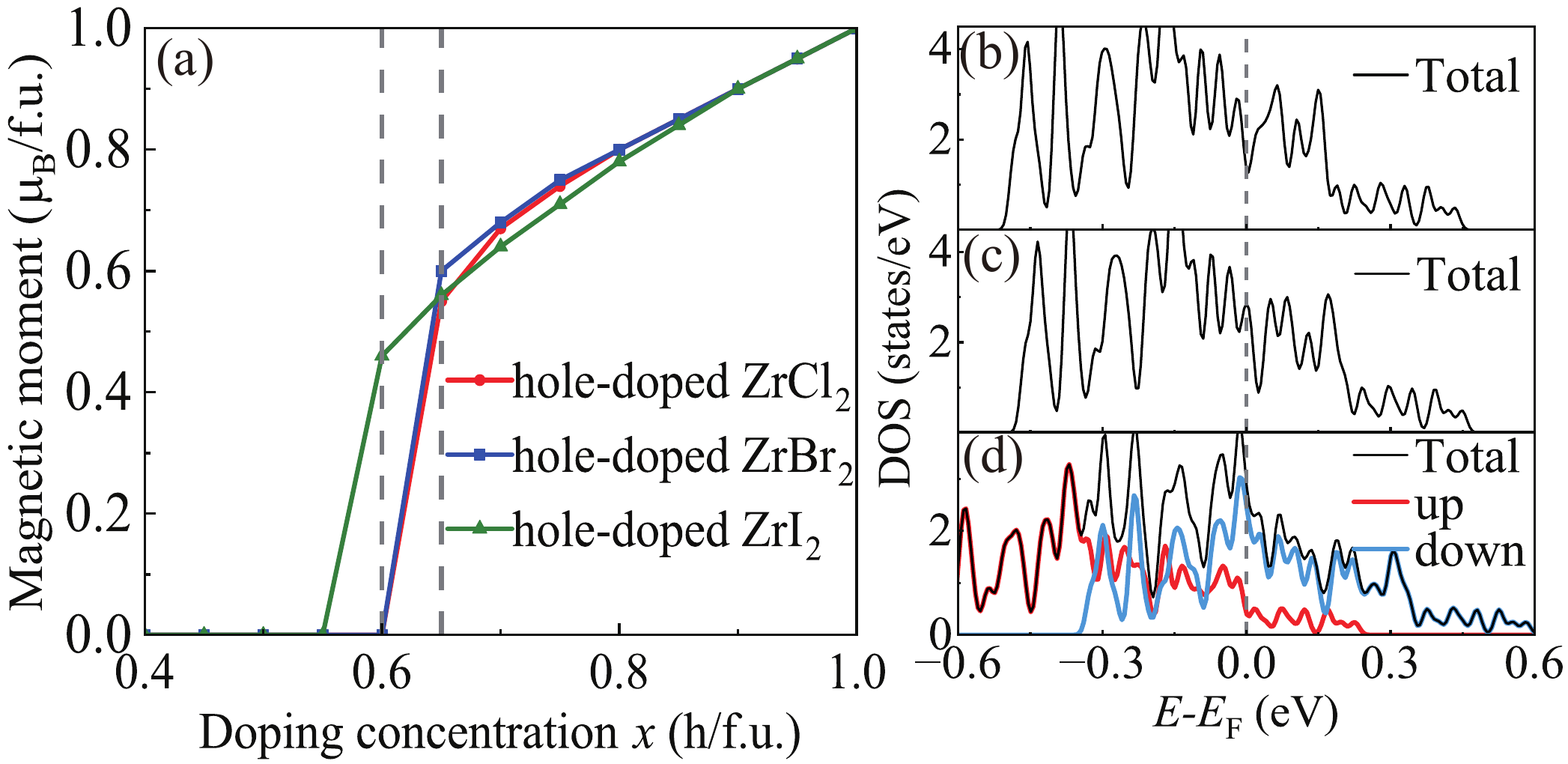}
	\caption{(a) Total magnetic moment ($\mu_{\text{B}}$/f.u.) of monolayer ZrX$_2$ (X = Cl, Br, I) as functions of hole doping concentration $x$. NM DOS of monolayer ZrI$_2$ at (b) $x = 0.55$ \textit{h}/f.u. and (c) $x = 0.60$ \textit{h}/f.u., FM DOS of monolayer ZrI$_2$ at (d) $x = 0.60$ \textit{h}/f.u.}
	\label{fig.5}
\end{figure}

At higher hole doping levels, the three ZrX$_2$ compounds develop spontaneous magnetization. Figure~\ref{fig.5}(a) shows the total magnetic moment as a function of hole concentration. Obviously, all three systems remain nonmagnetic up to the critical doping, after which the finite magnetic moments appear. The critical dopings are about 0.65~\textit{h}/f.u.\ for ZrCl$_2$ and ZrBr$_2$, and 0.6~\textit{h}/f.u.\ for ZrI$_2$, in agreement with previous reports for ZrCl$_2$~\cite{zrcl2mag34}. We also examined possible magnetic configurations (see the Supplemental Material Fig. S15) \cite{supplement}. For ZrI$_2$, the ferromagnetic state is stable across the relevant doping range, while for ZrCl$_2$ and ZrBr$_2$, increasing hole doping drives a transition from a weak stripe antiferromagnetic order to a ferromagnetic order.

The origin of this superconducting-magnetic transition can be traced to the electronic structure near the critical doping. Taking ZrI$_2$ as a representative case, Figs.~\ref{fig.5}(b),  ~\ref{fig.5}(c) and \ref{fig.5}(d) show the nonmagnetic DOS at 0.55~\textit{h}/f.u. and 0.6~\textit{h}/f.u., as well as the ferromagnetic (FM) DOS at 0.6~\textit{h}/f.u., respectively. As the hole concentration increases, the Fermi level shifts toward the DOS peak, producing a pronounced increase in $N(\varepsilon_\text{F})$ (Figs.~\ref{fig.5}(b) and  ~\ref{fig.5}(c)) \cite{supplement}. It fulfills the Stoner criterion $N(\varepsilon_\text{F})I > 1$ for magnetic instability~\cite{stoner1938}, where $N(\varepsilon_\text{F})$ is the DOS at the Fermi level and $I$ is the effective exchange interaction parameter. A similar mechanism occurs in LaBr$_2$, where a single excess electron per formula unit half-fills an interstitial band and gives intrinsic ferromagnetism with $T_{\text{Curie}} = 235$ K \cite{labr2prb35,labr2val36} (Table \ref{tab.1}). Thus, the magnetic instability arises from the progressive emptying of the interstitial band upon hole doping. When the band approaches half-filling, Stoner instability triggers ferromagnetism, accompanied by a dramatic decrease of $N(\varepsilon_\text{F})$ (from 2.86 $\times$ 2 states/eV (Figs.~\ref{fig.5}(c)) to 2.84 states/eV (Figs.~\ref{fig.5}(d))). Consequently, superconductivity is suppressed and eventually gives way to a magnetic state at higher doping.

\section{CONCLUSION}

In summary, monolayer ZrX$_2$ (X = Cl, Br, I) exhibits bipolar superconductivity, with electron doping yielding $T_c$ up to 30.0\,K at $x = 0.01$\,\textit{e}/f.u.\ and hole doping producing $\sim$2\ K above a threshold of $x = 0.2$--0.25\,\textit{h}/f.u. The asymmetry originates from the electride band structure, in which the flat conduction-band feature sits at the CBM, whereas the flat valence-band feature lies below the VBM. Specifically, the VBM is primarily composed of interstitial $d_{xy}/d_{x^2-y^2}$ states, which couple weakly to phonons. At elevated hole doping, Stoner instability drives the systems to magnetic state, yielding a phase diagram with competing superconducting and magnetic orders. These findings establish monolayer ZrX$_2$ as a platform for bipolar superconductivity and demonstrate that doping metallizes the flat band via Fermi-level shift, providing a strategy for engineering of high-$T_c$ superconductors.

\begin{acknowledgments}
We wish to thank Jun Fan for helpful discussions. This work was supported by the National Key R\&D Program of China (Grant No. 2022YFA1403103), the National Natural Science Foundation of China (Grant No. 12304167) and Young Talent of Lifting engineering for Science and Technology in Shandong, China.
\end{acknowledgments}

\normalem
\urlstyle{rm}
\bibliography{./Reference}

\begin{thebibliography}{57}%
\makeatletter
\providecommand \@ifxundefined [1]{%
 \@ifx{#1\undefined}
}%
\providecommand \@ifnum [1]{%
 \ifnum #1\expandafter \@firstoftwo
 \else \expandafter \@secondoftwo
 \fi
}%
\providecommand \@ifx [1]{%
 \ifx #1\expandafter \@firstoftwo
 \else \expandafter \@secondoftwo
 \fi
}%
\providecommand \natexlab [1]{#1}%
\providecommand \enquote  [1]{``#1''}%
\providecommand \bibnamefont  [1]{#1}%
\providecommand \bibfnamefont [1]{#1}%
\providecommand \citenamefont [1]{#1}%
\providecommand \href@noop [0]{\@secondoftwo}%
\providecommand \href [0]{\begingroup \@sanitize@url \@href}%
\providecommand \@href[1]{\@@startlink{#1}\@@href}%
\providecommand \@@href[1]{\endgroup#1\@@endlink}%
\providecommand \@sanitize@url [0]{\catcode `\\12\catcode `\$12\catcode
  `\&12\catcode `\#12\catcode `\^12\catcode `\_12\catcode `\%12\relax}%
\providecommand \@@startlink[1]{}%
\providecommand \@@endlink[0]{}%
\providecommand \url  [0]{\begingroup\@sanitize@url \@url }%
\providecommand \@url [1]{\endgroup\@href {#1}{\urlprefix }}%
\providecommand \urlprefix  [0]{URL }%
\providecommand \Eprint [0]{\href }%
\providecommand \doibase [0]{https://doi.org/}%
\providecommand \selectlanguage [0]{\@gobble}%
\providecommand \bibinfo  [0]{\@secondoftwo}%
\providecommand \bibfield  [0]{\@secondoftwo}%
\providecommand \translation [1]{[#1]}%
\providecommand \BibitemOpen [0]{}%
\providecommand \bibitemStop [0]{}%
\providecommand \bibitemNoStop [0]{.\EOS\space}%
\providecommand \EOS [0]{\spacefactor3000\relax}%
\providecommand \BibitemShut  [1]{\csname bibitem#1\endcsname}%
\let\auto@bib@innerbib\@empty
\bibitem [{\citenamefont {Torrance}\ \emph {et~al.}(1988)\citenamefont
  {Torrance}, \citenamefont {Tokura}, \citenamefont {Nazzal}, \citenamefont
  {Bezinge}, \citenamefont {Huang},\ and\ \citenamefont {Parkin}}]{cuprate1}%
  \BibitemOpen
  \bibfield  {author} {\bibinfo {author} {\bibfnamefont {J.~B.}\ \bibnamefont
  {Torrance}}, \bibinfo {author} {\bibfnamefont {Y.}~\bibnamefont {Tokura}},
  \bibinfo {author} {\bibfnamefont {A.~I.}\ \bibnamefont {Nazzal}}, \bibinfo
  {author} {\bibfnamefont {A.}~\bibnamefont {Bezinge}}, \bibinfo {author}
  {\bibfnamefont {T.~C.}\ \bibnamefont {Huang}},\ and\ \bibinfo {author}
  {\bibfnamefont {S.~S.~P.}\ \bibnamefont {Parkin}},\ }\bibfield  {title}
  {\bibinfo {title} {Anomalous disappearance of high-${T}_{c}$
  superconductivity at high hole concentration in metallic
  {La$_{2-x}$Sr$_{x}$CuO$_{4}$}},\ }\href
  {https://doi.org/10.1103/PhysRevLett.61.1127} {\bibfield  {journal} {\bibinfo
   {journal} {Phys. Rev. Lett.}\ }\textbf {\bibinfo {volume} {61}},\ \bibinfo
  {pages} {1127} (\bibinfo {year} {1988})}\BibitemShut {NoStop}%
\bibitem [{\citenamefont {Baiutti}\ \emph {et~al.}(2015)\citenamefont
  {Baiutti}, \citenamefont {Logvenov}, \citenamefont {Gregori}, \citenamefont
  {Cristiani}, \citenamefont {Wang}, \citenamefont {Sigle}, \citenamefont {van
  Aken},\ and\ \citenamefont {Maier}}]{cuprate2}%
  \BibitemOpen
  \bibfield  {author} {\bibinfo {author} {\bibfnamefont {F.}~\bibnamefont
  {Baiutti}}, \bibinfo {author} {\bibfnamefont {G.}~\bibnamefont {Logvenov}},
  \bibinfo {author} {\bibfnamefont {G.}~\bibnamefont {Gregori}}, \bibinfo
  {author} {\bibfnamefont {G.}~\bibnamefont {Cristiani}}, \bibinfo {author}
  {\bibfnamefont {Y.}~\bibnamefont {Wang}}, \bibinfo {author} {\bibfnamefont
  {W.}~\bibnamefont {Sigle}}, \bibinfo {author} {\bibfnamefont {P.~A.}\
  \bibnamefont {van Aken}},\ and\ \bibinfo {author} {\bibfnamefont
  {J.}~\bibnamefont {Maier}},\ }\bibfield  {title} {\bibinfo {title}
  {High-temperature superconductivity in space-charge regions of lanthanum
  cuprate induced by two-dimensional doping},\ }\href
  {https://doi.org/10.1038/ncomms9586} {\bibfield  {journal} {\bibinfo
  {journal} {Nat. Commun.}\ }\textbf {\bibinfo {volume} {6}},\ \bibinfo {pages}
  {8586} (\bibinfo {year} {2015})}\BibitemShut {NoStop}%
\bibitem [{\citenamefont {Lee}\ \emph {et~al.}(2006)\citenamefont {Lee},
  \citenamefont {Nagaosa},\ and\ \citenamefont {Wen}}]{cuprate3}%
  \BibitemOpen
  \bibfield  {author} {\bibinfo {author} {\bibfnamefont {P.~A.}\ \bibnamefont
  {Lee}}, \bibinfo {author} {\bibfnamefont {N.}~\bibnamefont {Nagaosa}},\ and\
  \bibinfo {author} {\bibfnamefont {X.-G.}\ \bibnamefont {Wen}},\ }\bibfield
  {title} {\bibinfo {title} {Doping a {Mott} insulator: Physics of
  high-temperature superconductivity},\ }\href
  {https://doi.org/10.1103/RevModPhys.78.17} {\bibfield  {journal} {\bibinfo
  {journal} {Rev. Mod. Phys.}\ }\textbf {\bibinfo {volume} {78}},\ \bibinfo
  {pages} {17} (\bibinfo {year} {2006})}\BibitemShut {NoStop}%
\bibitem [{\citenamefont {Wei}\ \emph {et~al.}(2016)\citenamefont {Wei},
  \citenamefont {Adamo}, \citenamefont {Nowadnick}, \citenamefont {Lochocki},
  \citenamefont {Chatterjee}, \citenamefont {Ruf}, \citenamefont {Beasley},
  \citenamefont {Schlom},\ and\ \citenamefont {Shen}}]{cuprate4}%
  \BibitemOpen
  \bibfield  {author} {\bibinfo {author} {\bibfnamefont {H.~I.}\ \bibnamefont
  {Wei}}, \bibinfo {author} {\bibfnamefont {C.}~\bibnamefont {Adamo}}, \bibinfo
  {author} {\bibfnamefont {E.~A.}\ \bibnamefont {Nowadnick}}, \bibinfo {author}
  {\bibfnamefont {E.~B.}\ \bibnamefont {Lochocki}}, \bibinfo {author}
  {\bibfnamefont {S.}~\bibnamefont {Chatterjee}}, \bibinfo {author}
  {\bibfnamefont {J.~P.}\ \bibnamefont {Ruf}}, \bibinfo {author} {\bibfnamefont
  {M.~R.}\ \bibnamefont {Beasley}}, \bibinfo {author} {\bibfnamefont {D.~G.}\
  \bibnamefont {Schlom}},\ and\ \bibinfo {author} {\bibfnamefont {K.~M.}\
  \bibnamefont {Shen}},\ }\bibfield  {title} {\bibinfo {title} {Electron doping
  of the parent cuprate {La$_{2}$CuO$_{4}$} without cation substitution},\
  }\href {https://doi.org/10.1103/PhysRevLett.117.147002} {\bibfield  {journal}
  {\bibinfo  {journal} {Phys. Rev. Lett.}\ }\textbf {\bibinfo {volume} {117}},\
  \bibinfo {pages} {147002} (\bibinfo {year} {2016})}\BibitemShut {NoStop}%
\bibitem [{\citenamefont {Mazin}(2010)}]{iron1}%
  \BibitemOpen
  \bibfield  {author} {\bibinfo {author} {\bibfnamefont {I.~I.}\ \bibnamefont
  {Mazin}},\ }\bibfield  {title} {\bibinfo {title} {Superconductivity gets an
  iron boost},\ }\href {https://doi.org/10.1038/nature08914} {\bibfield
  {journal} {\bibinfo  {journal} {Nature}\ }\textbf {\bibinfo {volume} {464}},\
  \bibinfo {pages} {183} (\bibinfo {year} {2010})}\BibitemShut {NoStop}%
\bibitem [{\citenamefont {Rotter}\ \emph {et~al.}(2008)\citenamefont {Rotter},
  \citenamefont {Tegel},\ and\ \citenamefont {Johrendt}}]{iron2}%
  \BibitemOpen
  \bibfield  {author} {\bibinfo {author} {\bibfnamefont {M.}~\bibnamefont
  {Rotter}}, \bibinfo {author} {\bibfnamefont {M.}~\bibnamefont {Tegel}},\ and\
  \bibinfo {author} {\bibfnamefont {D.}~\bibnamefont {Johrendt}},\ }\bibfield
  {title} {\bibinfo {title} {Superconductivity at 38 {K} in the iron arsenide
  {(Ba$_{1-x}$K$_{x}$)Fe$_{2}$As$_{2}$}},\ }\href
  {https://doi.org/10.1103/PhysRevLett.101.107006} {\bibfield  {journal}
  {\bibinfo  {journal} {Phys. Rev. Lett.}\ }\textbf {\bibinfo {volume} {101}},\
  \bibinfo {pages} {107006} (\bibinfo {year} {2008})}\BibitemShut {NoStop}%
\bibitem [{\citenamefont {Li}\ \emph {et~al.}(2009)\citenamefont {Li},
  \citenamefont {Luo}, \citenamefont {Wang}, \citenamefont {Chen},
  \citenamefont {Ren}, \citenamefont {Tao}, \citenamefont {Li}, \citenamefont
  {Lin}, \citenamefont {He}, \citenamefont {Zhu}, \citenamefont {Cao},\ and\
  \citenamefont {Xu}}]{iron3}%
  \BibitemOpen
  \bibfield  {author} {\bibinfo {author} {\bibfnamefont {L.~J.}\ \bibnamefont
  {Li}}, \bibinfo {author} {\bibfnamefont {Y.~K.}\ \bibnamefont {Luo}},
  \bibinfo {author} {\bibfnamefont {Q.~B.}\ \bibnamefont {Wang}}, \bibinfo
  {author} {\bibfnamefont {H.}~\bibnamefont {Chen}}, \bibinfo {author}
  {\bibfnamefont {Z.}~\bibnamefont {Ren}}, \bibinfo {author} {\bibfnamefont
  {Q.}~\bibnamefont {Tao}}, \bibinfo {author} {\bibfnamefont {Y.~K.}\
  \bibnamefont {Li}}, \bibinfo {author} {\bibfnamefont {X.}~\bibnamefont
  {Lin}}, \bibinfo {author} {\bibfnamefont {M.}~\bibnamefont {He}}, \bibinfo
  {author} {\bibfnamefont {Z.~W.}\ \bibnamefont {Zhu}}, \bibinfo {author}
  {\bibfnamefont {G.~H.}\ \bibnamefont {Cao}},\ and\ \bibinfo {author}
  {\bibfnamefont {Z.~A.}\ \bibnamefont {Xu}},\ }\bibfield  {title} {\bibinfo
  {title} {Superconductivity induced by {Ni} doping in {BaFe$_{2}$As$_{2}$}
  single crystals},\ }\href {https://doi.org/10.1088/1367-2630/11/2/025008}
  {\bibfield  {journal} {\bibinfo  {journal} {New J. Phys.}\ }\textbf {\bibinfo
  {volume} {11}},\ \bibinfo {pages} {025008} (\bibinfo {year}
  {2009})}\BibitemShut {NoStop}%
\bibitem [{\citenamefont {Kamihara}\ \emph {et~al.}(2008)\citenamefont
  {Kamihara}, \citenamefont {Watanabe}, \citenamefont {Hirano},\ and\
  \citenamefont {Hosono}}]{iron4}%
  \BibitemOpen
  \bibfield  {author} {\bibinfo {author} {\bibfnamefont {Y.}~\bibnamefont
  {Kamihara}}, \bibinfo {author} {\bibfnamefont {T.}~\bibnamefont {Watanabe}},
  \bibinfo {author} {\bibfnamefont {M.}~\bibnamefont {Hirano}},\ and\ \bibinfo
  {author} {\bibfnamefont {H.}~\bibnamefont {Hosono}},\ }\bibfield  {title}
  {\bibinfo {title} {Iron-based layered superconductor
  {La[O$_{1-x}$Fe$_{x}$]FeAs} (x = 0.05$-$0.12) with {$T_{c}$}= 26 {K}},\
  }\href {https://doi.org/10.1021/ja800073m} {\bibfield  {journal} {\bibinfo
  {journal} {J. Am. Chem. Soc.}\ }\textbf {\bibinfo {volume} {130}},\ \bibinfo
  {pages} {3296} (\bibinfo {year} {2008})}\BibitemShut {NoStop}%
\bibitem [{\citenamefont {Zhang}\ \emph {et~al.}(2020)\citenamefont {Zhang},
  \citenamefont {Xu}, \citenamefont {Zou}, \citenamefont {Ai}, \citenamefont
  {Dong}, \citenamefont {Huang}, \citenamefont {Leng}, \citenamefont {Liu},
  \citenamefont {Zhang}, \citenamefont {Jia} \emph {et~al.}}]{detector1}%
  \BibitemOpen
  \bibfield  {author} {\bibinfo {author} {\bibfnamefont {E.}~\bibnamefont
  {Zhang}}, \bibinfo {author} {\bibfnamefont {X.}~\bibnamefont {Xu}}, \bibinfo
  {author} {\bibfnamefont {Y.-C.}\ \bibnamefont {Zou}}, \bibinfo {author}
  {\bibfnamefont {L.}~\bibnamefont {Ai}}, \bibinfo {author} {\bibfnamefont
  {X.}~\bibnamefont {Dong}}, \bibinfo {author} {\bibfnamefont {C.}~\bibnamefont
  {Huang}}, \bibinfo {author} {\bibfnamefont {P.}~\bibnamefont {Leng}},
  \bibinfo {author} {\bibfnamefont {S.}~\bibnamefont {Liu}}, \bibinfo {author}
  {\bibfnamefont {Y.}~\bibnamefont {Zhang}}, \bibinfo {author} {\bibfnamefont
  {Z.}~\bibnamefont {Jia}}, \emph {et~al.},\ }\bibfield  {title} {\bibinfo
  {title} {Nonreciprocal superconducting {NbSe$_{2}$} antenna},\ }\href
  {https://doi.org/10.1038/s41467-020-19459-5} {\bibfield  {journal} {\bibinfo
  {journal} {Nat. Commun.}\ }\textbf {\bibinfo {volume} {11}},\ \bibinfo
  {pages} {5634} (\bibinfo {year} {2020})}\BibitemShut {NoStop}%
\bibitem [{\citenamefont {Han}\ \emph {et~al.}(2020)\citenamefont {Han},
  \citenamefont {Gao}, \citenamefont {Wu}, \citenamefont {Mu}, \citenamefont
  {Zhang}, \citenamefont {Mei}, \citenamefont {Chu}, \citenamefont {Xie},
  \citenamefont {Hu},\ and\ \citenamefont {Di}}]{flex2}%
  \BibitemOpen
  \bibfield  {author} {\bibinfo {author} {\bibfnamefont {X.}~\bibnamefont
  {Han}}, \bibinfo {author} {\bibfnamefont {M.}~\bibnamefont {Gao}}, \bibinfo
  {author} {\bibfnamefont {Y.}~\bibnamefont {Wu}}, \bibinfo {author}
  {\bibfnamefont {G.}~\bibnamefont {Mu}}, \bibinfo {author} {\bibfnamefont
  {M.}~\bibnamefont {Zhang}}, \bibinfo {author} {\bibfnamefont
  {Y.}~\bibnamefont {Mei}}, \bibinfo {author} {\bibfnamefont {P.~K.}\
  \bibnamefont {Chu}}, \bibinfo {author} {\bibfnamefont {X.}~\bibnamefont
  {Xie}}, \bibinfo {author} {\bibfnamefont {T.}~\bibnamefont {Hu}},\ and\
  \bibinfo {author} {\bibfnamefont {Z.}~\bibnamefont {Di}},\ }\bibfield
  {title} {\bibinfo {title} {Gate-tunable two-dimensional superconductivity
  revealed in flexible wafer-scale hybrid structures},\ }\href
  {https://doi.org/10.1039/D0TC04193A} {\bibfield  {journal} {\bibinfo
  {journal} {J. Mater. Chem. C}\ }\textbf {\bibinfo {volume} {8}},\ \bibinfo
  {pages} {14605} (\bibinfo {year} {2020})}\BibitemShut {NoStop}%
\bibitem [{\citenamefont {Zeng}\ \emph {et~al.}(2016)\citenamefont {Zeng},
  \citenamefont {Zhao}, \citenamefont {Li},\ and\ \citenamefont
  {Ni}}]{mos2bts}%
  \BibitemOpen
  \bibfield  {author} {\bibinfo {author} {\bibfnamefont {S.}~\bibnamefont
  {Zeng}}, \bibinfo {author} {\bibfnamefont {Y.}~\bibnamefont {Zhao}}, \bibinfo
  {author} {\bibfnamefont {G.}~\bibnamefont {Li}},\ and\ \bibinfo {author}
  {\bibfnamefont {J.}~\bibnamefont {Ni}},\ }\bibfield  {title} {\bibinfo
  {title} {Strongly enhanced superconductivity in doped monolayer {MoS$_{2}$}
  by strain},\ }\href {https://doi.org/10.1103/PhysRevB.94.024501} {\bibfield
  {journal} {\bibinfo  {journal} {Phys. Rev. B}\ }\textbf {\bibinfo {volume}
  {94}},\ \bibinfo {pages} {024501} (\bibinfo {year} {2016})}\BibitemShut
  {NoStop}%
\bibitem [{\citenamefont {Kong}\ \emph {et~al.}(2018)\citenamefont {Kong},
  \citenamefont {Gao}, \citenamefont {Yan}, \citenamefont {Lu},\ and\
  \citenamefont {Xiang}}]{asbts}%
  \BibitemOpen
  \bibfield  {author} {\bibinfo {author} {\bibfnamefont {X.}~\bibnamefont
  {Kong}}, \bibinfo {author} {\bibfnamefont {M.}~\bibnamefont {Gao}}, \bibinfo
  {author} {\bibfnamefont {X.-W.}\ \bibnamefont {Yan}}, \bibinfo {author}
  {\bibfnamefont {Z.-Y.}\ \bibnamefont {Lu}},\ and\ \bibinfo {author}
  {\bibfnamefont {T.}~\bibnamefont {Xiang}},\ }\bibfield  {title} {\bibinfo
  {title} {Superconductivity in electron-doped arsenene},\ }\href
  {https://doi.org/10.1088/1674-1056/27/4/046301} {\bibfield  {journal}
  {\bibinfo  {journal} {Chin. Phys. B}\ }\textbf {\bibinfo {volume} {27}},\
  \bibinfo {pages} {046301} (\bibinfo {year} {2018})}\BibitemShut {NoStop}%
\bibitem [{\citenamefont {Zhang}\ \emph {et~al.}(2018)\citenamefont {Zhang},
  \citenamefont {Fan}, \citenamefont {Wang}, \citenamefont {Zhang},
  \citenamefont {Wang}, \citenamefont {Li}, \citenamefont {He}, \citenamefont
  {Song}, \citenamefont {Ma},\ and\ \citenamefont {Xue}}]{snse2graphene23}%
  \BibitemOpen
  \bibfield  {author} {\bibinfo {author} {\bibfnamefont {Y.-M.}\ \bibnamefont
  {Zhang}}, \bibinfo {author} {\bibfnamefont {J.-Q.}\ \bibnamefont {Fan}},
  \bibinfo {author} {\bibfnamefont {W.-L.}\ \bibnamefont {Wang}}, \bibinfo
  {author} {\bibfnamefont {D.}~\bibnamefont {Zhang}}, \bibinfo {author}
  {\bibfnamefont {L.}~\bibnamefont {Wang}}, \bibinfo {author} {\bibfnamefont
  {W.}~\bibnamefont {Li}}, \bibinfo {author} {\bibfnamefont {K.}~\bibnamefont
  {He}}, \bibinfo {author} {\bibfnamefont {C.-L.}\ \bibnamefont {Song}},
  \bibinfo {author} {\bibfnamefont {X.-C.}\ \bibnamefont {Ma}},\ and\ \bibinfo
  {author} {\bibfnamefont {Q.-K.}\ \bibnamefont {Xue}},\ }\bibfield  {title}
  {\bibinfo {title} {Observation of interface superconductivity in a
  {SnSe$_{2}$}/epitaxial graphene van der {Waals} heterostructure},\ }\href
  {https://doi.org/10.1103/PhysRevB.98.220508} {\bibfield  {journal} {\bibinfo
  {journal} {Phys. Rev. B}\ }\textbf {\bibinfo {volume} {98}},\ \bibinfo
  {pages} {220508} (\bibinfo {year} {2018})}\BibitemShut {NoStop}%
\bibitem [{\citenamefont {Fan}\ \emph {et~al.}(2026)\citenamefont {Fan},
  \citenamefont {Qiu}, \citenamefont {Gong}, \citenamefont {Liu},\ and\
  \citenamefont {Lu}}]{snse2fan24}%
  \BibitemOpen
  \bibfield  {author} {\bibinfo {author} {\bibfnamefont {J.}~\bibnamefont
  {Fan}}, \bibinfo {author} {\bibfnamefont {X.-L.}\ \bibnamefont {Qiu}},
  \bibinfo {author} {\bibfnamefont {B.-C.}\ \bibnamefont {Gong}}, \bibinfo
  {author} {\bibfnamefont {K.}~\bibnamefont {Liu}},\ and\ \bibinfo {author}
  {\bibfnamefont {Z.-Y.}\ \bibnamefont {Lu}},\ }\bibfield  {title} {\bibinfo
  {title} {Interfacial superconductivity in the {type-III} heterostructure
  {SnSe$_{2}$}/{PtTe$_{2}$}},\ }\href
  {https://doi.org/10.1088/0256-307X/43/1/010711} {\bibfield  {journal}
  {\bibinfo  {journal} {Chin. Phys. Lett.}\ }\textbf {\bibinfo {volume} {43}},\
  \bibinfo {pages} {010711} (\bibinfo {year} {2026})}\BibitemShut {NoStop}%
\bibitem [{\citenamefont {Somoano}\ and\ \citenamefont
  {Rembaum}(1971)}]{limos2}%
  \BibitemOpen
  \bibfield  {author} {\bibinfo {author} {\bibfnamefont {R.~B.}\ \bibnamefont
  {Somoano}}\ and\ \bibinfo {author} {\bibfnamefont {A.}~\bibnamefont
  {Rembaum}},\ }\bibfield  {title} {\bibinfo {title} {Superconductivity in
  intercalated molybdenum disulfide},\ }\href
  {https://doi.org/10.1103/PhysRevLett.27.402} {\bibfield  {journal} {\bibinfo
  {journal} {Phys. Rev. Lett.}\ }\textbf {\bibinfo {volume} {27}},\ \bibinfo
  {pages} {402} (\bibinfo {year} {1971})}\BibitemShut {NoStop}%
\bibitem [{\citenamefont {Yamanaka}\ \emph {et~al.}(1996)\citenamefont
  {Yamanaka}, \citenamefont {Kawaji}, \citenamefont {Hotehama},\ and\
  \citenamefont {Ohashi}}]{lizrncl}%
  \BibitemOpen
  \bibfield  {author} {\bibinfo {author} {\bibfnamefont {S.}~\bibnamefont
  {Yamanaka}}, \bibinfo {author} {\bibfnamefont {H.}~\bibnamefont {Kawaji}},
  \bibinfo {author} {\bibfnamefont {K.-i.}\ \bibnamefont {Hotehama}},\ and\
  \bibinfo {author} {\bibfnamefont {M.}~\bibnamefont {Ohashi}},\ }\bibfield
  {title} {\bibinfo {title} {A new layer-structured nitride superconductor.
  lithium-intercalated $\beta$-zirconium nitride chloride, {Li$_x$ZrNCl}},\
  }\href {https://doi.org/https://doi.org/10.1002/adma.19960080917} {\bibfield
  {journal} {\bibinfo  {journal} {Adv. Mater.}\ }\textbf {\bibinfo {volume}
  {8}},\ \bibinfo {pages} {771} (\bibinfo {year} {1996})}\BibitemShut {NoStop}%
\bibitem [{\citenamefont {Chen}\ \emph {et~al.}(2022)\citenamefont {Chen},
  \citenamefont {Wang},\ and\ \citenamefont {Liu}}]{nainse}%
  \BibitemOpen
  \bibfield  {author} {\bibinfo {author} {\bibfnamefont {J.}~\bibnamefont
  {Chen}}, \bibinfo {author} {\bibfnamefont {X.}~\bibnamefont {Wang}},\ and\
  \bibinfo {author} {\bibfnamefont {X.}~\bibnamefont {Liu}},\ }\bibfield
  {title} {\bibinfo {title} {Emergence of superconductivity in an {InSe}
  monolayer: Roles of deposited metal and biaxial strain},\ }\href
  {https://doi.org/https://doi.org/10.1016/j.jpcs.2022.110823} {\bibfield
  {journal} {\bibinfo  {journal} {J. Phys. Chem. Solids}\ }\textbf {\bibinfo
  {volume} {168}},\ \bibinfo {pages} {110823} (\bibinfo {year}
  {2022})}\BibitemShut {NoStop}%
\bibitem [{\citenamefont {Chen}(2019)}]{inseelectron30}%
  \BibitemOpen
  \bibfield  {author} {\bibinfo {author} {\bibfnamefont {J.}~\bibnamefont
  {Chen}},\ }\bibfield  {title} {\bibinfo {title} {Phonon-mediated
  superconductivity in electron-doped monolayer {InSe}: a first-principles
  investigation},\ }\href
  {https://doi.org/https://doi.org/10.1016/j.jpcs.2018.09.039} {\bibfield
  {journal} {\bibinfo  {journal} {J. Phys. Chem. Solids}\ }\textbf {\bibinfo
  {volume} {125}},\ \bibinfo {pages} {23} (\bibinfo {year} {2019})}\BibitemShut
  {NoStop}%
\bibitem [{\citenamefont {Shao}\ \emph {et~al.}(2014)\citenamefont {Shao},
  \citenamefont {Lu}, \citenamefont {Lv},\ and\ \citenamefont
  {Sun}}]{phosphorene29}%
  \BibitemOpen
  \bibfield  {author} {\bibinfo {author} {\bibfnamefont {D.}~\bibnamefont
  {Shao}}, \bibinfo {author} {\bibfnamefont {W.}~\bibnamefont {Lu}}, \bibinfo
  {author} {\bibfnamefont {H.}~\bibnamefont {Lv}},\ and\ \bibinfo {author}
  {\bibfnamefont {Y.}~\bibnamefont {Sun}},\ }\bibfield  {title} {\bibinfo
  {title} {Electron-doped phosphorene: A potential monolayer superconductor},\
  }\href {https://doi.org/10.1209/0295-5075/108/67004} {\bibfield  {journal}
  {\bibinfo  {journal} {EPL-Europhys. Lett}\ }\textbf {\bibinfo {volume}
  {108}},\ \bibinfo {pages} {67004} (\bibinfo {year} {2014})}\BibitemShut
  {NoStop}%
\bibitem [{\citenamefont {Banerjee}\ \emph {et~al.}(2021)\citenamefont
  {Banerjee}, \citenamefont {Barone},\ and\ \citenamefont {Picozzi}}]{magadd1}%
  \BibitemOpen
  \bibfield  {author} {\bibinfo {author} {\bibfnamefont {H.}~\bibnamefont
  {Banerjee}}, \bibinfo {author} {\bibfnamefont {P.}~\bibnamefont {Barone}},\
  and\ \bibinfo {author} {\bibfnamefont {S.}~\bibnamefont {Picozzi}},\
  }\bibfield  {title} {\bibinfo {title} {Half-metallic ferromagnetism in
  layered {CdOHCl} induced by hole doping},\ }\href
  {https://doi.org/10.1088/2053-1583/abe4ba} {\bibfield  {journal} {\bibinfo
  {journal} {2D Mater.}\ }\textbf {\bibinfo {volume} {8}},\ \bibinfo {pages}
  {025027} (\bibinfo {year} {2021})}\BibitemShut {NoStop}%
\bibitem [{\citenamefont {Liu}\ \emph {et~al.}(2021)\citenamefont {Liu},
  \citenamefont {Wang}, \citenamefont {Jia}, \citenamefont {Liu}, \citenamefont
  {Yin}, \citenamefont {Guan},\ and\ \citenamefont {Cheng}}]{magadd2}%
  \BibitemOpen
  \bibfield  {author} {\bibinfo {author} {\bibfnamefont {C.}~\bibnamefont
  {Liu}}, \bibinfo {author} {\bibfnamefont {B.}~\bibnamefont {Wang}}, \bibinfo
  {author} {\bibfnamefont {G.}~\bibnamefont {Jia}}, \bibinfo {author}
  {\bibfnamefont {P.}~\bibnamefont {Liu}}, \bibinfo {author} {\bibfnamefont
  {H.}~\bibnamefont {Yin}}, \bibinfo {author} {\bibfnamefont {S.}~\bibnamefont
  {Guan}},\ and\ \bibinfo {author} {\bibfnamefont {Z.}~\bibnamefont {Cheng}},\
  }\bibfield  {title} {\bibinfo {title} {Tunable magnetism in ferroelectric
  $\alpha$-{In$_{2}$Se$_{3}$} by hole-doping},\ }\href
  {https://doi.org/10.1063/5.0039842} {\bibfield  {journal} {\bibinfo
  {journal} {Appl. Phys. Lett.}\ }\textbf {\bibinfo {volume} {118}},\ \bibinfo
  {pages} {072902} (\bibinfo {year} {2021})}\BibitemShut {NoStop}%
\bibitem [{\citenamefont {Iordanidou}\ \emph {et~al.}(2020)\citenamefont
  {Iordanidou}, \citenamefont {Houssa},\ and\ \citenamefont
  {Persson}}]{magadd3}%
  \BibitemOpen
  \bibfield  {author} {\bibinfo {author} {\bibfnamefont {K.}~\bibnamefont
  {Iordanidou}}, \bibinfo {author} {\bibfnamefont {M.}~\bibnamefont {Houssa}},\
  and\ \bibinfo {author} {\bibfnamefont {C.}~\bibnamefont {Persson}},\
  }\bibfield  {title} {\bibinfo {title} {Carrier-mediated ferromagnetism in
  two-dimensional {PtS$_{2}$}},\ }\href {https://doi.org/10.1039/C9RA09756B}
  {\bibfield  {journal} {\bibinfo  {journal} {RSC Adv.}\ }\textbf {\bibinfo
  {volume} {10}},\ \bibinfo {pages} {952} (\bibinfo {year} {2020})}\BibitemShut
  {NoStop}%
\bibitem [{\citenamefont {He}\ \emph {et~al.}(2022)\citenamefont {He},
  \citenamefont {Chen}, \citenamefont {Wang}, \citenamefont {Yang},
  \citenamefont {Yang}, \citenamefont {Shen}, \citenamefont {Xu}, \citenamefont
  {Jiang}, \citenamefont {Chai}, \citenamefont {Wong} \emph
  {et~al.}}]{zrcl2mag34}%
  \BibitemOpen
  \bibfield  {author} {\bibinfo {author} {\bibfnamefont {J.}~\bibnamefont
  {He}}, \bibinfo {author} {\bibfnamefont {Y.}~\bibnamefont {Chen}}, \bibinfo
  {author} {\bibfnamefont {Z.}~\bibnamefont {Wang}}, \bibinfo {author}
  {\bibfnamefont {M.}~\bibnamefont {Yang}}, \bibinfo {author} {\bibfnamefont
  {T.}~\bibnamefont {Yang}}, \bibinfo {author} {\bibfnamefont {L.}~\bibnamefont
  {Shen}}, \bibinfo {author} {\bibfnamefont {X.}~\bibnamefont {Xu}}, \bibinfo
  {author} {\bibfnamefont {Y.}~\bibnamefont {Jiang}}, \bibinfo {author}
  {\bibfnamefont {J.}~\bibnamefont {Chai}}, \bibinfo {author} {\bibfnamefont
  {L.~M.}\ \bibnamefont {Wong}}, \emph {et~al.},\ }\bibfield  {title} {\bibinfo
  {title} {Formation of magnetic anionic electrons by hole doping},\ }\href
  {https://doi.org/10.1039/D2TC00564F} {\bibfield  {journal} {\bibinfo
  {journal} {J. Mater. Chem. C}\ }\textbf {\bibinfo {volume} {10}},\ \bibinfo
  {pages} {7674} (\bibinfo {year} {2022})}\BibitemShut {NoStop}%
\bibitem [{\citenamefont {Alidoosti}\ \emph {et~al.}(2021)\citenamefont
  {Alidoosti}, \citenamefont {Esfahani},\ and\ \citenamefont
  {Asgari}}]{insehole31}%
  \BibitemOpen
  \bibfield  {author} {\bibinfo {author} {\bibfnamefont {M.}~\bibnamefont
  {Alidoosti}}, \bibinfo {author} {\bibfnamefont {D.~N.}\ \bibnamefont
  {Esfahani}},\ and\ \bibinfo {author} {\bibfnamefont {R.}~\bibnamefont
  {Asgari}},\ }\bibfield  {title} {\bibinfo {title} {Charge density wave and
  superconducting phase in monolayer {InSe}},\ }\href
  {https://doi.org/10.1103/PhysRevB.103.035411} {\bibfield  {journal} {\bibinfo
   {journal} {Phys. Rev. B}\ }\textbf {\bibinfo {volume} {103}},\ \bibinfo
  {pages} {035411} (\bibinfo {year} {2021})}\BibitemShut {NoStop}%
\bibitem [{\citenamefont {An}\ and\ \citenamefont {Pickett}(2001)}]{sigmabond}%
  \BibitemOpen
  \bibfield  {author} {\bibinfo {author} {\bibfnamefont {J.~M.}\ \bibnamefont
  {An}}\ and\ \bibinfo {author} {\bibfnamefont {W.~E.}\ \bibnamefont
  {Pickett}},\ }\bibfield  {title} {\bibinfo {title} {Superconductivity of
  {MgB$_2$}: Covalent bonds driven metallic},\ }\href
  {https://doi.org/10.1103/PhysRevLett.86.4366} {\bibfield  {journal} {\bibinfo
   {journal} {Phys. Rev. Lett.}\ }\textbf {\bibinfo {volume} {86}},\ \bibinfo
  {pages} {4366} (\bibinfo {year} {2001})}\BibitemShut {NoStop}%
\bibitem [{\citenamefont {Gao}\ \emph {et~al.}(2015)\citenamefont {Gao},
  \citenamefont {Lu},\ and\ \citenamefont {Xiang}}]{cnphysics}%
  \BibitemOpen
  \bibfield  {author} {\bibinfo {author} {\bibfnamefont {M.}~\bibnamefont
  {Gao}}, \bibinfo {author} {\bibfnamefont {Z.-Y.}\ \bibnamefont {Lu}},\ and\
  \bibinfo {author} {\bibfnamefont {T.}~\bibnamefont {Xiang}},\ }\bibfield
  {title} {\bibinfo {title} {Finding high-temperature superconductors by
  metallizing the $\sigma$-bonding electrons},\ }\href
  {https://doi.org/10.7693/wl20150701} {\bibfield  {journal} {\bibinfo
  {journal} {PHYSICS}\ }\textbf {\bibinfo {volume} {44}},\ \bibinfo {pages}
  {421} (\bibinfo {year} {2015})}\BibitemShut {NoStop}%
\bibitem [{\citenamefont {Zhou}\ \emph {et~al.}(2024)\citenamefont {Zhou},
  \citenamefont {You}, \citenamefont {Zhao}, \citenamefont {Feng},\ and\
  \citenamefont {Shen}}]{van_ene3}%
  \BibitemOpen
  \bibfield  {author} {\bibinfo {author} {\bibfnamefont {J.}~\bibnamefont
  {Zhou}}, \bibinfo {author} {\bibfnamefont {J.-Y.}\ \bibnamefont {You}},
  \bibinfo {author} {\bibfnamefont {Y.-M.}\ \bibnamefont {Zhao}}, \bibinfo
  {author} {\bibfnamefont {Y.~P.}\ \bibnamefont {Feng}},\ and\ \bibinfo
  {author} {\bibfnamefont {L.}~\bibnamefont {Shen}},\ }\bibfield  {title}
  {\bibinfo {title} {Van der {Waals} electrides},\ }\href
  {https://doi.org/10.1021/acs.accounts.4c00394} {\bibfield  {journal}
  {\bibinfo  {journal} {Acc. Chem. Res.}\ }\textbf {\bibinfo {volume} {57}},\
  \bibinfo {pages} {2572} (\bibinfo {year} {2024})}\BibitemShut {NoStop}%
\bibitem [{\citenamefont {Zeng}\ \emph {et~al.}(2018)\citenamefont {Zeng},
  \citenamefont {Zhao}, \citenamefont {Li},\ and\ \citenamefont
  {Yang}}]{ca2ncal6}%
  \BibitemOpen
  \bibfield  {author} {\bibinfo {author} {\bibfnamefont {X.}~\bibnamefont
  {Zeng}}, \bibinfo {author} {\bibfnamefont {S.}~\bibnamefont {Zhao}}, \bibinfo
  {author} {\bibfnamefont {Z.}~\bibnamefont {Li}},\ and\ \bibinfo {author}
  {\bibfnamefont {J.}~\bibnamefont {Yang}},\ }\bibfield  {title} {\bibinfo
  {title} {Electron-phonon interaction in a {Ca$_{2}$N} monolayer: Intrinsic
  mobility of electrene},\ }\href {https://doi.org/10.1103/PhysRevB.98.155443}
  {\bibfield  {journal} {\bibinfo  {journal} {Phys. Rev. B}\ }\textbf {\bibinfo
  {volume} {98}},\ \bibinfo {pages} {155443} (\bibinfo {year}
  {2018})}\BibitemShut {NoStop}%
\bibitem [{\citenamefont {Qiu}\ \emph {et~al.}(2022)\citenamefont {Qiu},
  \citenamefont {Zhang}, \citenamefont {Yang}, \citenamefont {Lu},\ and\
  \citenamefont {Liu}}]{ba2ncal13}%
  \BibitemOpen
  \bibfield  {author} {\bibinfo {author} {\bibfnamefont {X.-L.}\ \bibnamefont
  {Qiu}}, \bibinfo {author} {\bibfnamefont {J.-F.}\ \bibnamefont {Zhang}},
  \bibinfo {author} {\bibfnamefont {H.-C.}\ \bibnamefont {Yang}}, \bibinfo
  {author} {\bibfnamefont {Z.-Y.}\ \bibnamefont {Lu}},\ and\ \bibinfo {author}
  {\bibfnamefont {K.}~\bibnamefont {Liu}},\ }\bibfield  {title} {\bibinfo
  {title} {Superconductivity in monolayer {Ba$_{2}$N} electride:
  First-principles study},\ }\href
  {https://doi.org/10.1103/PhysRevB.105.165101} {\bibfield  {journal} {\bibinfo
   {journal} {Phys. Rev. B}\ }\textbf {\bibinfo {volume} {105}},\ \bibinfo
  {pages} {165101} (\bibinfo {year} {2022})}\BibitemShut {NoStop}%
\bibitem [{\citenamefont {Zhou}\ \emph {et~al.}(2020)\citenamefont {Zhou},
  \citenamefont {Feng},\ and\ \citenamefont {Shen}}]{labr2prb35}%
  \BibitemOpen
  \bibfield  {author} {\bibinfo {author} {\bibfnamefont {J.}~\bibnamefont
  {Zhou}}, \bibinfo {author} {\bibfnamefont {Y.~P.}\ \bibnamefont {Feng}},\
  and\ \bibinfo {author} {\bibfnamefont {L.}~\bibnamefont {Shen}},\ }\bibfield
  {title} {\bibinfo {title} {Atomic-orbital-free intrinsic ferromagnetism in
  electrenes},\ }\href {https://doi.org/10.1103/PhysRevB.102.180407} {\bibfield
   {journal} {\bibinfo  {journal} {Phys. Rev. B}\ }\textbf {\bibinfo {volume}
  {102}},\ \bibinfo {pages} {180407} (\bibinfo {year} {2020})}\BibitemShut
  {NoStop}%
\bibitem [{\citenamefont {Zhao}\ \emph {et~al.}(2019)\citenamefont {Zhao},
  \citenamefont {Ma}, \citenamefont {Lei}, \citenamefont {Wang}, \citenamefont
  {Huang},\ and\ \citenamefont {Dai}}]{labr2val36}%
  \BibitemOpen
  \bibfield  {author} {\bibinfo {author} {\bibfnamefont {P.}~\bibnamefont
  {Zhao}}, \bibinfo {author} {\bibfnamefont {Y.}~\bibnamefont {Ma}}, \bibinfo
  {author} {\bibfnamefont {C.}~\bibnamefont {Lei}}, \bibinfo {author}
  {\bibfnamefont {H.}~\bibnamefont {Wang}}, \bibinfo {author} {\bibfnamefont
  {B.}~\bibnamefont {Huang}},\ and\ \bibinfo {author} {\bibfnamefont
  {Y.}~\bibnamefont {Dai}},\ }\bibfield  {title} {\bibinfo {title}
  {Single-layer {LaBr$_{2}$}: Two-dimensional valleytronic semiconductor with
  spontaneous spin and valley polarizations},\ }\href
  {https://doi.org/10.1063/1.5129311} {\bibfield  {journal} {\bibinfo
  {journal} {Appl. Phys. Lett.}\ }\textbf {\bibinfo {volume} {115}},\ \bibinfo
  {pages} {261605} (\bibinfo {year} {2019})}\BibitemShut {NoStop}%
\bibitem [{\citenamefont {Yi}\ \emph {et~al.}(2025)\citenamefont {Yi},
  \citenamefont {Li}, \citenamefont {Cao}, \citenamefont {Han},\ and\
  \citenamefont {Zhang}}]{zrx2cal61}%
  \BibitemOpen
  \bibfield  {author} {\bibinfo {author} {\bibfnamefont {Y.}~\bibnamefont
  {Yi}}, \bibinfo {author} {\bibfnamefont {Z.}~\bibnamefont {Li}}, \bibinfo
  {author} {\bibfnamefont {S.}~\bibnamefont {Cao}}, \bibinfo {author}
  {\bibfnamefont {J.}~\bibnamefont {Han}},\ and\ \bibinfo {author}
  {\bibfnamefont {Z.}~\bibnamefont {Zhang}},\ }\bibfield  {title} {\bibinfo
  {title} {Geometrical stability, electrical contact and optical properties for
  {ZrX$_{2}$ (X = Cl, Br, I)}/{Zr$_{2}$Cl$_{2}$} semiconductor-metal
  heterojunctions},\ }\href
  {https://doi.org/https://doi.org/10.1016/j.apsusc.2024.161730} {\bibfield
  {journal} {\bibinfo  {journal} {Appl. Surf. Sci.}\ }\textbf {\bibinfo
  {volume} {682}},\ \bibinfo {pages} {161730} (\bibinfo {year}
  {2025})}\BibitemShut {NoStop}%
\bibitem [{\citenamefont {Guo}\ \emph {et~al.}(2021)\citenamefont {Guo},
  \citenamefont {Lu}, \citenamefont {Wang}, \citenamefont {Zhao}, \citenamefont
  {Hu}, \citenamefont {Yuan},\ and\ \citenamefont {Ren}}]{2zrx2}%
  \BibitemOpen
  \bibfield  {author} {\bibinfo {author} {\bibfnamefont {J.}~\bibnamefont
  {Guo}}, \bibinfo {author} {\bibfnamefont {Z.}~\bibnamefont {Lu}}, \bibinfo
  {author} {\bibfnamefont {K.}~\bibnamefont {Wang}}, \bibinfo {author}
  {\bibfnamefont {X.}~\bibnamefont {Zhao}}, \bibinfo {author} {\bibfnamefont
  {G.}~\bibnamefont {Hu}}, \bibinfo {author} {\bibfnamefont {X.}~\bibnamefont
  {Yuan}},\ and\ \bibinfo {author} {\bibfnamefont {J.}~\bibnamefont {Ren}},\
  }\bibfield  {title} {\bibinfo {title} {Large valley polarization in a novel
  two-dimensional semiconductor {H-ZrX$_{2}$ (X = Cl, Br, I)}},\ }\href
  {https://doi.org/10.1088/1361-648X/ac394f} {\bibfield  {journal} {\bibinfo
  {journal} {J. Phys.: Condens. Matter}\ }\textbf {\bibinfo {volume} {34}},\
  \bibinfo {pages} {075701} (\bibinfo {year} {2021})}\BibitemShut {NoStop}%
\bibitem [{\citenamefont {Wen}\ \emph {et~al.}(2023)\citenamefont {Wen},
  \citenamefont {Peng}, \citenamefont {Zhang},\ and\ \citenamefont
  {Wang}}]{thelezri239}%
  \BibitemOpen
  \bibfield  {author} {\bibinfo {author} {\bibfnamefont {J.}~\bibnamefont
  {Wen}}, \bibinfo {author} {\bibfnamefont {J.}~\bibnamefont {Peng}}, \bibinfo
  {author} {\bibfnamefont {B.}~\bibnamefont {Zhang}},\ and\ \bibinfo {author}
  {\bibfnamefont {Z.-Y.}\ \bibnamefont {Wang}},\ }\bibfield  {title} {\bibinfo
  {title} {Theoretical determination of superior high-temperature
  thermoelectricity in an n-type doped {2H-ZrI$_{2}$} monolayer},\ }\href
  {https://doi.org/10.1039/D2NR06054J} {\bibfield  {journal} {\bibinfo
  {journal} {Nanoscale}\ }\textbf {\bibinfo {volume} {15}},\ \bibinfo {pages}
  {4397} (\bibinfo {year} {2023})}\BibitemShut {NoStop}%
\bibitem [{\citenamefont {Chen}\ \emph {et~al.}(2023)\citenamefont {Chen},
  \citenamefont {Guo}, \citenamefont {Fan}, \citenamefont {Li}, \citenamefont
  {Wu}, \citenamefont {Ma}, \citenamefont {Gong}, \citenamefont {Dong},
  \citenamefont {Wang}, \citenamefont {Xu},\ and\ \citenamefont
  {An}}]{nanozri240}%
  \BibitemOpen
  \bibfield  {author} {\bibinfo {author} {\bibfnamefont {J.}~\bibnamefont
  {Chen}}, \bibinfo {author} {\bibfnamefont {Y.}~\bibnamefont {Guo}}, \bibinfo
  {author} {\bibfnamefont {X.}~\bibnamefont {Fan}}, \bibinfo {author}
  {\bibfnamefont {Y.}~\bibnamefont {Li}}, \bibinfo {author} {\bibfnamefont
  {Y.}~\bibnamefont {Wu}}, \bibinfo {author} {\bibfnamefont {C.}~\bibnamefont
  {Ma}}, \bibinfo {author} {\bibfnamefont {S.}~\bibnamefont {Gong}}, \bibinfo
  {author} {\bibfnamefont {X.}~\bibnamefont {Dong}}, \bibinfo {author}
  {\bibfnamefont {T.}~\bibnamefont {Wang}}, \bibinfo {author} {\bibfnamefont
  {G.}~\bibnamefont {Xu}},\ and\ \bibinfo {author} {\bibfnamefont
  {Y.}~\bibnamefont {An}},\ }\bibfield  {title} {\bibinfo {title} {Nanodevices
  from and electronic transport properties of {ZrI$_{2}$} monolayers},\ }\href
  {https://doi.org/10.1103/PhysRevApplied.20.064048} {\bibfield  {journal}
  {\bibinfo  {journal} {Phys. Rev. Appl.}\ }\textbf {\bibinfo {volume} {20}},\
  \bibinfo {pages} {064048} (\bibinfo {year} {2023})}\BibitemShut {NoStop}%
\bibitem [{\citenamefont {Giannozzi}\ \emph {et~al.}(2009)\citenamefont
  {Giannozzi}, \citenamefont {Baroni}, \citenamefont {Bonini}, \citenamefont
  {Calandra}, \citenamefont {Car}, \citenamefont {Cavazzoni}, \citenamefont
  {Ceresoli}, \citenamefont {Chiarotti}, \citenamefont {Cococcioni},
  \citenamefont {Dabo} \emph {et~al.}}]{qe44}%
  \BibitemOpen
  \bibfield  {author} {\bibinfo {author} {\bibfnamefont {P.}~\bibnamefont
  {Giannozzi}}, \bibinfo {author} {\bibfnamefont {S.}~\bibnamefont {Baroni}},
  \bibinfo {author} {\bibfnamefont {N.}~\bibnamefont {Bonini}}, \bibinfo
  {author} {\bibfnamefont {M.}~\bibnamefont {Calandra}}, \bibinfo {author}
  {\bibfnamefont {R.}~\bibnamefont {Car}}, \bibinfo {author} {\bibfnamefont
  {C.}~\bibnamefont {Cavazzoni}}, \bibinfo {author} {\bibfnamefont
  {D.}~\bibnamefont {Ceresoli}}, \bibinfo {author} {\bibfnamefont {G.~L.}\
  \bibnamefont {Chiarotti}}, \bibinfo {author} {\bibfnamefont {M.}~\bibnamefont
  {Cococcioni}}, \bibinfo {author} {\bibfnamefont {I.}~\bibnamefont {Dabo}},
  \emph {et~al.},\ }\bibfield  {title} {\bibinfo {title} {{Quantum ESPRESSO}: a
  modular and open-source software project for quantum simulations of
  materials},\ }\href {https://doi.org/10.1088/0953-8984/21/39/395502}
  {\bibfield  {journal} {\bibinfo  {journal} {J. Phys.: Condens. Matter}\
  }\textbf {\bibinfo {volume} {21}},\ \bibinfo {pages} {395502} (\bibinfo
  {year} {2009})}\BibitemShut {NoStop}%
\bibitem [{\citenamefont {Giannozzi}\ \emph {et~al.}(2017)\citenamefont
  {Giannozzi}, \citenamefont {Andreussi}, \citenamefont {Brumme}, \citenamefont
  {Bunau}, \citenamefont {Buongiorno~Nardelli}, \citenamefont {Calandra},
  \citenamefont {Car}, \citenamefont {Cavazzoni}, \citenamefont {Ceresoli},
  \citenamefont {Cococcioni}, \citenamefont {Colonna}, \citenamefont
  {Carnimeo}, \citenamefont {Dal~Corso} \emph {et~al.}}]{qe45}%
  \BibitemOpen
  \bibfield  {author} {\bibinfo {author} {\bibfnamefont {P.}~\bibnamefont
  {Giannozzi}}, \bibinfo {author} {\bibfnamefont {O.}~\bibnamefont
  {Andreussi}}, \bibinfo {author} {\bibfnamefont {T.}~\bibnamefont {Brumme}},
  \bibinfo {author} {\bibfnamefont {O.}~\bibnamefont {Bunau}}, \bibinfo
  {author} {\bibfnamefont {M.}~\bibnamefont {Buongiorno~Nardelli}}, \bibinfo
  {author} {\bibfnamefont {M.}~\bibnamefont {Calandra}}, \bibinfo {author}
  {\bibfnamefont {R.}~\bibnamefont {Car}}, \bibinfo {author} {\bibfnamefont
  {C.}~\bibnamefont {Cavazzoni}}, \bibinfo {author} {\bibfnamefont
  {D.}~\bibnamefont {Ceresoli}}, \bibinfo {author} {\bibfnamefont
  {M.}~\bibnamefont {Cococcioni}}, \bibinfo {author} {\bibfnamefont
  {N.}~\bibnamefont {Colonna}}, \bibinfo {author} {\bibfnamefont
  {I.}~\bibnamefont {Carnimeo}}, \bibinfo {author} {\bibfnamefont
  {A.}~\bibnamefont {Dal~Corso}}, \emph {et~al.},\ }\bibfield  {title}
  {\bibinfo {title} {Advanced capabilities for materials modelling with
  {Quantum ESPRESSO}},\ }\href {https://doi.org/10.1088/1361-648X/aa8f79}
  {\bibfield  {journal} {\bibinfo  {journal} {J. Phys.: Condens. Matter}\
  }\textbf {\bibinfo {volume} {29}},\ \bibinfo {pages} {465901} (\bibinfo
  {year} {2017})}\BibitemShut {NoStop}%
\bibitem [{\citenamefont {Giannozzi}\ \emph {et~al.}(2020)\citenamefont
  {Giannozzi}, \citenamefont {Baseggio}, \citenamefont {Bonfà}, \citenamefont
  {Brunato}, \citenamefont {Car}, \citenamefont {Carnimeo}, \citenamefont
  {Cavazzoni}, \citenamefont {de~Gironcoli}, \citenamefont {Delugas},
  \citenamefont {Ferrari~Ruffino}, \citenamefont {Ferretti}, \citenamefont
  {Marzari}, \citenamefont {Timrov}, \citenamefont {Urru},\ and\ \citenamefont
  {Baroni}}]{qe46}%
  \BibitemOpen
  \bibfield  {author} {\bibinfo {author} {\bibfnamefont {P.}~\bibnamefont
  {Giannozzi}}, \bibinfo {author} {\bibfnamefont {O.}~\bibnamefont {Baseggio}},
  \bibinfo {author} {\bibfnamefont {P.}~\bibnamefont {Bonfà}}, \bibinfo
  {author} {\bibfnamefont {D.}~\bibnamefont {Brunato}}, \bibinfo {author}
  {\bibfnamefont {R.}~\bibnamefont {Car}}, \bibinfo {author} {\bibfnamefont
  {I.}~\bibnamefont {Carnimeo}}, \bibinfo {author} {\bibfnamefont
  {C.}~\bibnamefont {Cavazzoni}}, \bibinfo {author} {\bibfnamefont
  {S.}~\bibnamefont {de~Gironcoli}}, \bibinfo {author} {\bibfnamefont
  {P.}~\bibnamefont {Delugas}}, \bibinfo {author} {\bibfnamefont
  {F.}~\bibnamefont {Ferrari~Ruffino}}, \bibinfo {author} {\bibfnamefont
  {A.}~\bibnamefont {Ferretti}}, \bibinfo {author} {\bibfnamefont
  {N.}~\bibnamefont {Marzari}}, \bibinfo {author} {\bibfnamefont
  {I.}~\bibnamefont {Timrov}}, \bibinfo {author} {\bibfnamefont
  {A.}~\bibnamefont {Urru}},\ and\ \bibinfo {author} {\bibfnamefont
  {S.}~\bibnamefont {Baroni}},\ }\bibfield  {title} {\bibinfo {title} {{Quantum
  ESPRESSO} toward the exascale},\ }\href {https://doi.org/10.1063/5.0005082}
  {\bibfield  {journal} {\bibinfo  {journal} {J. Chem. Phys.}\ }\textbf
  {\bibinfo {volume} {152}},\ \bibinfo {pages} {154105} (\bibinfo {year}
  {2020})}\BibitemShut {NoStop}%
\bibitem [{\citenamefont {Bl\"ochl}(1994)}]{paw41}%
  \BibitemOpen
  \bibfield  {author} {\bibinfo {author} {\bibfnamefont {P.~E.}\ \bibnamefont
  {Bl\"ochl}},\ }\bibfield  {title} {\bibinfo {title} {Projector augmented-wave
  method},\ }\href {https://doi.org/10.1103/PhysRevB.50.17953} {\bibfield
  {journal} {\bibinfo  {journal} {Phys. Rev. B}\ }\textbf {\bibinfo {volume}
  {50}},\ \bibinfo {pages} {17953} (\bibinfo {year} {1994})}\BibitemShut
  {NoStop}%
\bibitem [{\citenamefont {Kresse}\ and\ \citenamefont {Joubert}(1999)}]{paw42}%
  \BibitemOpen
  \bibfield  {author} {\bibinfo {author} {\bibfnamefont {G.}~\bibnamefont
  {Kresse}}\ and\ \bibinfo {author} {\bibfnamefont {D.}~\bibnamefont
  {Joubert}},\ }\bibfield  {title} {\bibinfo {title} {From ultrasoft
  pseudopotentials to the projector augmented-wave method},\ }\href
  {https://doi.org/10.1103/PhysRevB.59.1758} {\bibfield  {journal} {\bibinfo
  {journal} {Phys. Rev. B}\ }\textbf {\bibinfo {volume} {59}},\ \bibinfo
  {pages} {1758} (\bibinfo {year} {1999})}\BibitemShut {NoStop}%
\bibitem [{\citenamefont {Perdew}\ \emph {et~al.}(1996)\citenamefont {Perdew},
  \citenamefont {Burke},\ and\ \citenamefont {Ernzerhof}}]{gga43}%
  \BibitemOpen
  \bibfield  {author} {\bibinfo {author} {\bibfnamefont {J.~P.}\ \bibnamefont
  {Perdew}}, \bibinfo {author} {\bibfnamefont {K.}~\bibnamefont {Burke}},\ and\
  \bibinfo {author} {\bibfnamefont {M.}~\bibnamefont {Ernzerhof}},\ }\bibfield
  {title} {\bibinfo {title} {{Generalized Gradient Approximation} made
  simple},\ }\href {https://doi.org/10.1103/PhysRevLett.77.3865} {\bibfield
  {journal} {\bibinfo  {journal} {Phys. Rev. Lett.}\ }\textbf {\bibinfo
  {volume} {77}},\ \bibinfo {pages} {3865} (\bibinfo {year}
  {1996})}\BibitemShut {NoStop}%
\bibitem [{\citenamefont {Baroni}\ \emph {et~al.}(2001)\citenamefont {Baroni},
  \citenamefont {de~Gironcoli}, \citenamefont {Dal~Corso},\ and\ \citenamefont
  {Giannozzi}}]{dfpt47}%
  \BibitemOpen
  \bibfield  {author} {\bibinfo {author} {\bibfnamefont {S.}~\bibnamefont
  {Baroni}}, \bibinfo {author} {\bibfnamefont {S.}~\bibnamefont
  {de~Gironcoli}}, \bibinfo {author} {\bibfnamefont {A.}~\bibnamefont
  {Dal~Corso}},\ and\ \bibinfo {author} {\bibfnamefont {P.}~\bibnamefont
  {Giannozzi}},\ }\bibfield  {title} {\bibinfo {title} {Phonons and related
  crystal properties from density-functional perturbation theory},\ }\href
  {https://doi.org/10.1103/RevModPhys.73.515} {\bibfield  {journal} {\bibinfo
  {journal} {Rev. Mod. Phys.}\ }\textbf {\bibinfo {volume} {73}},\ \bibinfo
  {pages} {515} (\bibinfo {year} {2001})}\BibitemShut {NoStop}%
\bibitem [{\citenamefont {Giustino}(2017)}]{dfpt48}%
  \BibitemOpen
  \bibfield  {author} {\bibinfo {author} {\bibfnamefont {F.}~\bibnamefont
  {Giustino}},\ }\bibfield  {title} {\bibinfo {title} {Electron-phonon
  interactions from first principles},\ }\href
  {https://doi.org/10.1103/RevModPhys.89.015003} {\bibfield  {journal}
  {\bibinfo  {journal} {Rev. Mod. Phys.}\ }\textbf {\bibinfo {volume} {89}},\
  \bibinfo {pages} {015003} (\bibinfo {year} {2017})}\BibitemShut {NoStop}%
\bibitem [{\citenamefont {Allen}(1972)}]{allen49}%
  \BibitemOpen
  \bibfield  {author} {\bibinfo {author} {\bibfnamefont {P.~B.}\ \bibnamefont
  {Allen}},\ }\bibfield  {title} {\bibinfo {title} {Neutron spectroscopy of
  superconductors},\ }\href {https://doi.org/10.1103/PhysRevB.6.2577}
  {\bibfield  {journal} {\bibinfo  {journal} {Phys. Rev. B}\ }\textbf {\bibinfo
  {volume} {6}},\ \bibinfo {pages} {2577} (\bibinfo {year} {1972})}\BibitemShut
  {NoStop}%
\bibitem [{\citenamefont {Lee}\ \emph {et~al.}(1995)\citenamefont {Lee},
  \citenamefont {Chang},\ and\ \citenamefont {Cohen}}]{miu50}%
  \BibitemOpen
  \bibfield  {author} {\bibinfo {author} {\bibfnamefont {K.-H.}\ \bibnamefont
  {Lee}}, \bibinfo {author} {\bibfnamefont {K.~J.}\ \bibnamefont {Chang}},\
  and\ \bibinfo {author} {\bibfnamefont {M.~L.}\ \bibnamefont {Cohen}},\
  }\bibfield  {title} {\bibinfo {title} {First-principles calculations of the
  {Coulomb} pseudopotential $\mu^{*}$: Application to {Al}},\ }\href
  {https://doi.org/10.1103/PhysRevB.52.1425} {\bibfield  {journal} {\bibinfo
  {journal} {Phys. Rev. B}\ }\textbf {\bibinfo {volume} {52}},\ \bibinfo
  {pages} {1425} (\bibinfo {year} {1995})}\BibitemShut {NoStop}%
\bibitem [{\citenamefont {Richardson}\ and\ \citenamefont
  {Ashcroft}(1997)}]{miu51}%
  \BibitemOpen
  \bibfield  {author} {\bibinfo {author} {\bibfnamefont {C.~F.}\ \bibnamefont
  {Richardson}}\ and\ \bibinfo {author} {\bibfnamefont {N.~W.}\ \bibnamefont
  {Ashcroft}},\ }\bibfield  {title} {\bibinfo {title} {High temperature
  superconductivity in metallic hydrogen: Electron-electron enhancements},\
  }\href {https://doi.org/10.1103/PhysRevLett.78.118} {\bibfield  {journal}
  {\bibinfo  {journal} {Phys. Rev. Lett.}\ }\textbf {\bibinfo {volume} {78}},\
  \bibinfo {pages} {118} (\bibinfo {year} {1997})}\BibitemShut {NoStop}%
\bibitem [{sup()}]{supplement}%
  \BibitemOpen
  \href@noop {} {}\bibinfo {note} {See {Supplemental Materials} at
  \url{https://} for the convergence tests, structural characterization and
  stability of ZrX$_2$ monolayers (the stucture of bulk ZrX$_2$, phonon
  dispersion, {AIMD}, {ELF}, band stuctures of monolayer ZrX$_2$), and
  electron-doped superconductivity of ZrBr$_2$ and ZrI$_2$ (band stuctures,
  PDOS and EPC properties at optimal doping concentrations), as well as
  hole-doped superconductivity and magnetism of ZrBr$_2$ and ZrI$_2$ (the
  evolutions of band structures, PDOS and Fermi surface around the
  superconducting critical points).}\BibitemShut {Stop}%
\bibitem [{\citenamefont {Giustino}\ \emph {et~al.}(2007)\citenamefont
  {Giustino}, \citenamefont {Cohen},\ and\ \citenamefont {Louie}}]{epw1}%
  \BibitemOpen
  \bibfield  {author} {\bibinfo {author} {\bibfnamefont {F.}~\bibnamefont
  {Giustino}}, \bibinfo {author} {\bibfnamefont {M.~L.}\ \bibnamefont
  {Cohen}},\ and\ \bibinfo {author} {\bibfnamefont {S.~G.}\ \bibnamefont
  {Louie}},\ }\bibfield  {title} {\bibinfo {title} {Electron-phonon interaction
  using {Wannier} functions},\ }\href
  {https://doi.org/10.1103/PhysRevB.76.165108} {\bibfield  {journal} {\bibinfo
  {journal} {Phys. Rev. B}\ }\textbf {\bibinfo {volume} {76}},\ \bibinfo
  {pages} {165108} (\bibinfo {year} {2007})}\BibitemShut {NoStop}%
\bibitem [{\citenamefont {Noffsinger}\ \emph {et~al.}(2010)\citenamefont
  {Noffsinger}, \citenamefont {Giustino}, \citenamefont {Malone}, \citenamefont
  {Park}, \citenamefont {Louie},\ and\ \citenamefont {Cohen}}]{epw2}%
  \BibitemOpen
  \bibfield  {author} {\bibinfo {author} {\bibfnamefont {J.}~\bibnamefont
  {Noffsinger}}, \bibinfo {author} {\bibfnamefont {F.}~\bibnamefont
  {Giustino}}, \bibinfo {author} {\bibfnamefont {B.~D.}\ \bibnamefont
  {Malone}}, \bibinfo {author} {\bibfnamefont {C.-H.}\ \bibnamefont {Park}},
  \bibinfo {author} {\bibfnamefont {S.~G.}\ \bibnamefont {Louie}},\ and\
  \bibinfo {author} {\bibfnamefont {M.~L.}\ \bibnamefont {Cohen}},\ }\bibfield
  {title} {\bibinfo {title} {{EPW}: A program for calculating the
  electron–phonon coupling using maximally localized {Wannier} functions},\
  }\href {https://doi.org/10.1016/j.cpc.2010.08.027} {\bibfield  {journal}
  {\bibinfo  {journal} {Comput. Phys. Commun.}\ }\textbf {\bibinfo {volume}
  {181}},\ \bibinfo {pages} {2140} (\bibinfo {year} {2010})}\BibitemShut
  {NoStop}%
\bibitem [{\citenamefont {Poncé}\ \emph {et~al.}(2016)\citenamefont {Poncé},
  \citenamefont {Margine}, \citenamefont {Verdi},\ and\ \citenamefont
  {Giustino}}]{epw3}%
  \BibitemOpen
  \bibfield  {author} {\bibinfo {author} {\bibfnamefont {S.}~\bibnamefont
  {Poncé}}, \bibinfo {author} {\bibfnamefont {E.}~\bibnamefont {Margine}},
  \bibinfo {author} {\bibfnamefont {C.}~\bibnamefont {Verdi}},\ and\ \bibinfo
  {author} {\bibfnamefont {F.}~\bibnamefont {Giustino}},\ }\bibfield  {title}
  {\bibinfo {title} {{EPW}: Electron–phonon coupling, transport and
  superconducting properties using maximally localized {Wannier} functions},\
  }\href {https://doi.org/10.1016/j.cpc.2016.07.028} {\bibfield  {journal}
  {\bibinfo  {journal} {Comput. Phys. Commun.}\ }\textbf {\bibinfo {volume}
  {209}},\ \bibinfo {pages} {116} (\bibinfo {year} {2016})}\BibitemShut
  {NoStop}%
\bibitem [{\citenamefont {Eda}\ \emph {et~al.}(2011)\citenamefont {Eda},
  \citenamefont {Yamaguchi}, \citenamefont {Voiry}, \citenamefont {Fujita},
  \citenamefont {Chen},\ and\ \citenamefont {Chhowalla}}]{mos2}%
  \BibitemOpen
  \bibfield  {author} {\bibinfo {author} {\bibfnamefont {G.}~\bibnamefont
  {Eda}}, \bibinfo {author} {\bibfnamefont {H.}~\bibnamefont {Yamaguchi}},
  \bibinfo {author} {\bibfnamefont {D.}~\bibnamefont {Voiry}}, \bibinfo
  {author} {\bibfnamefont {T.}~\bibnamefont {Fujita}}, \bibinfo {author}
  {\bibfnamefont {M.}~\bibnamefont {Chen}},\ and\ \bibinfo {author}
  {\bibfnamefont {M.}~\bibnamefont {Chhowalla}},\ }\bibfield  {title} {\bibinfo
  {title} {Photoluminescence from chemically exfoliated {MoS$_{2}$}},\ }\href
  {https://doi.org/10.1021/nl201874w} {\bibfield  {journal} {\bibinfo
  {journal} {Nano Lett.}\ }\textbf {\bibinfo {volume} {11}},\ \bibinfo {pages}
  {5111} (\bibinfo {year} {2011})}\BibitemShut {NoStop}%
\bibitem [{\citenamefont {Cisar}\ \emph {et~al.}(1979)\citenamefont {Cisar},
  \citenamefont {Corbett},\ and\ \citenamefont {Daake}}]{zrclexp60}%
  \BibitemOpen
  \bibfield  {author} {\bibinfo {author} {\bibfnamefont {A.}~\bibnamefont
  {Cisar}}, \bibinfo {author} {\bibfnamefont {J.~D.}\ \bibnamefont {Corbett}},\
  and\ \bibinfo {author} {\bibfnamefont {R.~L.}\ \bibnamefont {Daake}},\
  }\bibfield  {title} {\bibinfo {title} {The zirconium dichloride phase region.
  synthesis, structure, and photoelectron spectral studies of {3R-ZrCl$_{2}$},
  {6T-Zr$_{1.05}$Cl$_{2}$}, and related phases},\ }\href
  {https://doi.org/10.1021/ic50193a060} {\bibfield  {journal} {\bibinfo
  {journal} {Inorg. Chem.}\ }\textbf {\bibinfo {volume} {18}},\ \bibinfo
  {pages} {836} (\bibinfo {year} {1979})}\BibitemShut {NoStop}%
\bibitem [{\citenamefont {Rasche}\ \emph {et~al.}(2022)\citenamefont {Rasche},
  \citenamefont {Brunner}, \citenamefont {Schramm}, \citenamefont {Ghimire},
  \citenamefont {Nitzsche}, \citenamefont {B{\"u}chner}, \citenamefont
  {Giraud}, \citenamefont {Richter},\ and\ \citenamefont
  {Dufouleur}}]{cleagra64}%
  \BibitemOpen
  \bibfield  {author} {\bibinfo {author} {\bibfnamefont {B.}~\bibnamefont
  {Rasche}}, \bibinfo {author} {\bibfnamefont {J.}~\bibnamefont {Brunner}},
  \bibinfo {author} {\bibfnamefont {T.}~\bibnamefont {Schramm}}, \bibinfo
  {author} {\bibfnamefont {M.~P.}\ \bibnamefont {Ghimire}}, \bibinfo {author}
  {\bibfnamefont {U.}~\bibnamefont {Nitzsche}}, \bibinfo {author}
  {\bibfnamefont {B.}~\bibnamefont {B{\"u}chner}}, \bibinfo {author}
  {\bibfnamefont {R.}~\bibnamefont {Giraud}}, \bibinfo {author} {\bibfnamefont
  {M.}~\bibnamefont {Richter}},\ and\ \bibinfo {author} {\bibfnamefont
  {J.}~\bibnamefont {Dufouleur}},\ }\bibfield  {title} {\bibinfo {title}
  {Determination of cleavage energy and efficient nanostructuring of layered
  materials by atomic force microscopy},\ }\href
  {https://doi.org/10.1021/acs.nanolett.1c04868} {\bibfield  {journal}
  {\bibinfo  {journal} {Nano Lett.}\ }\textbf {\bibinfo {volume} {22}},\
  \bibinfo {pages} {3550} (\bibinfo {year} {2022})}\BibitemShut {NoStop}%
\bibitem [{\citenamefont {Wang}\ \emph {et~al.}(2015)\citenamefont {Wang},
  \citenamefont {Dai}, \citenamefont {Li}, \citenamefont {Yang}, \citenamefont
  {Srolovitz},\ and\ \citenamefont {Zheng}}]{cleagra65}%
  \BibitemOpen
  \bibfield  {author} {\bibinfo {author} {\bibfnamefont {W.}~\bibnamefont
  {Wang}}, \bibinfo {author} {\bibfnamefont {S.}~\bibnamefont {Dai}}, \bibinfo
  {author} {\bibfnamefont {X.}~\bibnamefont {Li}}, \bibinfo {author}
  {\bibfnamefont {J.}~\bibnamefont {Yang}}, \bibinfo {author} {\bibfnamefont
  {D.~J.}\ \bibnamefont {Srolovitz}},\ and\ \bibinfo {author} {\bibfnamefont
  {Q.}~\bibnamefont {Zheng}},\ }\bibfield  {title} {\bibinfo {title}
  {Measurement of the cleavage energy of graphite},\ }\href
  {https://doi.org/10.1038/ncomms8853} {\bibfield  {journal} {\bibinfo
  {journal} {Nat. Commun.}\ }\textbf {\bibinfo {volume} {6}},\ \bibinfo {pages}
  {7853} (\bibinfo {year} {2015})}\BibitemShut {NoStop}%
\bibitem [{\citenamefont {Allen}\ and\ \citenamefont {Dynes}(1975)}]{ad}%
  \BibitemOpen
  \bibfield  {author} {\bibinfo {author} {\bibfnamefont {P.~B.}\ \bibnamefont
  {Allen}}\ and\ \bibinfo {author} {\bibfnamefont {R.~C.}\ \bibnamefont
  {Dynes}},\ }\bibfield  {title} {\bibinfo {title} {Transition temperature of
  strong-coupled superconductors reanalyzed},\ }\href
  {https://doi.org/10.1103/PhysRevB.12.905} {\bibfield  {journal} {\bibinfo
  {journal} {Phys. Rev. B}\ }\textbf {\bibinfo {volume} {12}},\ \bibinfo
  {pages} {905} (\bibinfo {year} {1975})}\BibitemShut {NoStop}%
\bibitem [{\citenamefont {Gu}\ \emph {et~al.}(2026)\citenamefont {Gu},
  \citenamefont {Hu}, \citenamefont {Jiang},\ and\ \citenamefont
  {Xiang}}]{gu2026}%
  \BibitemOpen
  \bibfield  {author} {\bibinfo {author} {\bibfnamefont {Y.}~\bibnamefont
  {Gu}}, \bibinfo {author} {\bibfnamefont {J.}~\bibnamefont {Hu}}, \bibinfo
  {author} {\bibfnamefont {H.}~\bibnamefont {Jiang}},\ and\ \bibinfo {author}
  {\bibfnamefont {T.}~\bibnamefont {Xiang}},\ }\bibfield  {title} {\bibinfo
  {title} {Optimal coloring and strain-enhanced superconductivity in
  {Li$_{n}$B$_{n+1}$C$_{n-1}$}},\ }\href
  {https://doi.org/10.1038/s42005-026-02495-w} {\bibfield  {journal} {\bibinfo
  {journal} {Commun. Phys.}\ }\textbf {\bibinfo {volume} {9}},\ \bibinfo
  {pages} {81} (\bibinfo {year} {2026})}\BibitemShut {NoStop}%
\bibitem [{\citenamefont {Stoner}(1938)}]{stoner1938}%
  \BibitemOpen
  \bibfield  {author} {\bibinfo {author} {\bibfnamefont {E.~C.}\ \bibnamefont
  {Stoner}},\ }\bibfield  {title} {\bibinfo {title} {Collective electron
  ferromagnetism},\ }\href {https://doi.org/10.1098/rspa.1938.0066} {\bibfield
  {journal} {\bibinfo  {journal} {Proc. R. Soc. London, Ser. A}\ }\textbf
  {\bibinfo {volume} {165}},\ \bibinfo {pages} {372} (\bibinfo {year}
  {1938})}\BibitemShut {NoStop}%
\end{thebibliography}%


\begin{thebibliography}{4}%
\makeatletter
\providecommand \@ifxundefined [1]{%
 \@ifx{#1\undefined}
}%
\providecommand \@ifnum [1]{%
 \ifnum #1\expandafter \@firstoftwo
 \else \expandafter \@secondoftwo
 \fi
}%
\providecommand \@ifx [1]{%
 \ifx #1\expandafter \@firstoftwo
 \else \expandafter \@secondoftwo
 \fi
}%
\providecommand \natexlab [1]{#1}%
\providecommand \enquote  [1]{``#1''}%
\providecommand \bibnamefont  [1]{#1}%
\providecommand \bibfnamefont [1]{#1}%
\providecommand \citenamefont [1]{#1}%
\providecommand \href@noop [0]{\@secondoftwo}%
\providecommand \href [0]{\begingroup \@sanitize@url \@href}%
\providecommand \@href[1]{\@@startlink{#1}\@@href}%
\providecommand \@@href[1]{\endgroup#1\@@endlink}%
\providecommand \@sanitize@url [0]{\catcode `\\12\catcode `\$12\catcode
  `\&12\catcode `\#12\catcode `\^12\catcode `\_12\catcode `\%12\relax}%
\providecommand \@@startlink[1]{}%
\providecommand \@@endlink[0]{}%
\providecommand \url  [0]{\begingroup\@sanitize@url \@url }%
\providecommand \@url [1]{\endgroup\@href {#1}{\urlprefix }}%
\providecommand \urlprefix  [0]{URL }%
\providecommand \Eprint [0]{\href }%
\providecommand \doibase [0]{https://doi.org/}%
\providecommand \selectlanguage [0]{\@gobble}%
\providecommand \bibinfo  [0]{\@secondoftwo}%
\providecommand \bibfield  [0]{\@secondoftwo}%
\providecommand \translation [1]{[#1]}%
\providecommand \BibitemOpen [0]{}%
\providecommand \bibitemStop [0]{}%
\providecommand \bibitemNoStop [0]{.\EOS\space}%
\providecommand \EOS [0]{\spacefactor3000\relax}%
\providecommand \BibitemShut  [1]{\csname bibitem#1\endcsname}%
\let\auto@bib@innerbib\@empty
\bibitem [{\citenamefont {Ahmed}\ \textit{et}~\textit{al.}(2025)\citenamefont
  {Ahmed}, \citenamefont {Gonz{\'a}lez-Cataldo}, \citenamefont {Robinson},\
  and\ \citenamefont {Militzer}}]{electride1}%
  \BibitemOpen
  \bibfield  {author} {\bibinfo {author} {\bibfnamefont {S.}~\bibnamefont
  {Ahmed}}, \bibinfo {author} {\bibfnamefont {F.}~\bibnamefont
  {Gonz{\'a}lez-Cataldo}}, \bibinfo {author} {\bibfnamefont {V.~N.}\
  \bibnamefont {Robinson}},\ and\ \bibinfo {author} {\bibfnamefont
  {B.}~\bibnamefont {Militzer}},\ }\href {https://doi.org/10.1103/14dp-bf1n}
  {\bibfield  {journal} {\bibinfo  {journal} {Phys. Rev. B}\ }\textbf {\bibinfo
  {volume} {112}},\ \bibinfo {pages} {165111} (\bibinfo {year}
  {2025})}\BibitemShut {NoStop}%
\bibitem [{\citenamefont {Weaver}\ \textit{et}~\textit{al.}(2023)\citenamefont
  {Weaver}, \citenamefont {Sundberg}, \citenamefont {Slamowitz}, \citenamefont
  {Radomsky},\ and\ \citenamefont {Lanetti}}]{badelf}%
  \BibitemOpen
  \bibfield  {author} {\bibinfo {author} {\bibfnamefont {S.~M.}\ \bibnamefont
  {Weaver}}, \bibinfo {author} {\bibfnamefont {J.~D.}\ \bibnamefont
  {Sundberg}}, \bibinfo {author} {\bibfnamefont {C.~C.}\ \bibnamefont
  {Slamowitz}}, \bibinfo {author} {\bibfnamefont {R.~C.}\ \bibnamefont
  {Radomsky}},\ and\ \bibinfo {author} {\bibfnamefont {M.~G.}\ \bibnamefont
  {Lanetti}},\ }\href {https://doi.org/10.1021/jacs.3c10876} {\bibfield
  {journal} {\bibinfo  {journal} {J. Am. Chem. Soc.}\ }\textbf {\bibinfo
  {volume} {145}},\ \bibinfo {pages} {26472} (\bibinfo {year}
  {2023})}\BibitemShut {NoStop}%
\bibitem [{\citenamefont {de-la Roza}\
  \textit{et}~\textit{al.}(2014)\citenamefont {de-la Roza}, \citenamefont
  {Johnson},\ and\ \citenamefont {Luaña}}]{cc1}%
  \BibitemOpen
  \bibfield  {author} {\bibinfo {author} {\bibfnamefont {A.~O.}\ \bibnamefont
  {de-la Roza}}, \bibinfo {author} {\bibfnamefont {E.~R.}\ \bibnamefont
  {Johnson}},\ and\ \bibinfo {author} {\bibfnamefont {V.}~\bibnamefont
  {Luaña}},\ }\href {https://doi.org/10.1016/j.cpc.2013.10.026} {\bibfield
  {journal} {\bibinfo  {journal} {Comput. Phys. Commun.}\ }\textbf {\bibinfo
  {volume} {185}},\ \bibinfo {pages} {1007} (\bibinfo {year}
  {2014})}\BibitemShut {NoStop}%
\bibitem [{\citenamefont {de-la Roza}\
  \textit{et}~\textit{al.}(2009)\citenamefont {de-la Roza}, \citenamefont
  {Blanco}, \citenamefont {Pendás},\ and\ \citenamefont {Luaña}}]{cc2}%
  \BibitemOpen
  \bibfield  {author} {\bibinfo {author} {\bibfnamefont {A.~O.}\ \bibnamefont
  {de-la Roza}}, \bibinfo {author} {\bibfnamefont {M.}~\bibnamefont {Blanco}},
  \bibinfo {author} {\bibfnamefont {A.~M.}\ \bibnamefont {Pendás}},\ and\
  \bibinfo {author} {\bibfnamefont {V.}~\bibnamefont {Luaña}},\ }\href
  {https://doi.org/10.1016/j.cpc.2008.07.018} {\bibfield  {journal} {\bibinfo
  {journal} {Comput. Phys. Commun.}\ }\textbf {\bibinfo {volume} {180}},\
  \bibinfo {pages} {157} (\bibinfo {year} {2009})}\BibitemShut {NoStop}%
\end{thebibliography}%

\end{document}


\title{Electron- and hole-doped superconductivity in two-dimensional electride ZrX$_2$ (X = Cl, Br, I)}
\author{Tian-Xin Liu}
\affiliation{Department of Physics, School of Science, Jiangsu University of Science and Technology, Zhenjiang 212100, China}
\author{Xin-Yang Li}
\affiliation{Department of Physics, School of Science, Jiangsu University of Science and Technology, Zhenjiang 212100, China}
\author{Zuo-Wei Yin}
\affiliation{Department of Physics, School of Science, Jiangsu University of Science and Technology, Zhenjiang 212100, China}
\author{Jing Jiang}
\affiliation{School of Physics and Key Laboratory of Quantum State Construction and Manipulation (Ministry of Education), Renmin University of China, Beijing 100872, China}
\author{Ben-Chao Gong}\email{bcgong@just.edu.cn}
\affiliation{Department of Physics, School of Science, Jiangsu University of Science and Technology, Zhenjiang 212100, China}
\author{Wei Fan}
\affiliation{Department of Physics, School of Science, Jiangsu University of Science and Technology, Zhenjiang 212100, China}
\author{Xiao-Le Qiu}\email{qiu.xiaole@163.com}
\affiliation{School of Physics and Electronic Information, Weifang University, Weifang 261061, China}
\author{Kai Liu}
\affiliation{School of Physics and Key Laboratory of Quantum State Construction and Manipulation (Ministry of Education), Renmin University of China, Beijing 100872, China}

\maketitle
\renewcommand{\figurename}{FIG.S}
\renewcommand{\tablename}{TABLE S}

\onecolumngrid

\section*{Supplemental Material}

\subsection*{1. The convergence tests}

\begin{figure}[htbp]
	\centering
	\includegraphics[width=0.9\columnwidth]{./pics1.pdf}
	\caption{Convergence tests for monolayer ZrCl$_2$. (a) Total energy versus charge-density cutoff (ecutrho). (b) Total energy versus kinetic-energy cutoff (ecutwfc). (c) Total energy versus Gaussian smearing width. Superconducting Tc of optimally electron-doped (d) ZrCl$_2$, (e) ZrBr$_2$, and (f) ZrI$_2$, as functions of the el\_ph\_sigma parameter. The dashed lines mark the respective convergence thresholds.}
	\label{fig.s1}
\end{figure}

\newpage

\begin{figure}[htbp]
	\centering
	\includegraphics[width=0.8\columnwidth]{./pics2.pdf}
	\caption{The q-mesh convergence tests. (a) The superconducting $T_c$ as functions of q-mesh calculated by the direct QE/DFPT method. (b) The $T_c$ and $\lambda$ as functions of q-mesh calculated by the EPW codes. All results are for ZrCl$_2$ at 0.05 \textit{e}/f.u. with $\mu^{*}$ = 0.1, using the McMillan-Allen-Dynes formula.}
	\label{fig.s2}
\end{figure}

\subsection*{2. Structural characterization and stability of ZrX$_2$ monolayers}

\begin{figure}[htb]
	\centering
	\includegraphics[width=0.75\columnwidth]{./pics3.pdf}
	\caption{(a) Side view of the bulk ZrX$_2$ (X = Cl, Br, I). (b) Cleavage energy of ZrX$_2$ (X = Cl, Br, I). (c) Phonon dispersion of monolayer ZrCl$_2$. (d) Phonon dispersion of monolayer ZrCl$_2$ at a doping concentration $x = 0.35$ \textit{e}/f.u.}
	\label{fig.s3}
\end{figure}

\begin{figure}[htb]
	\centering
	\includegraphics[width=0.8\columnwidth]{./pics4.pdf}
	\caption{Band structure and density of states (a), and phonon dispersion (b) of monolayer ZrBr$_2$. Corresponding results for monolayer ZrI$_2$ are shown in (c) and (d), respectively.}
	\label{fig.s4}
\end{figure}

\begin{figure}[htb]
	\centering
	\includegraphics[width=1.0\columnwidth]{./pics5.pdf}
	\caption{Time evolution of total energy for monolayer ZrX$_2$ (X = Cl, Br, I) at 300 K in \textit{ab initio} molecular dynamics simulation. The insets show the top (left) and side (right) views of the atomic structure in a snapshot at 6 \textit{p}s.}
	\label{fig.s5}
\end{figure}

\clearpage

\begin{figure}[htb]
	\centering
	\includegraphics[width=0.9\columnwidth]{./pics6.pdf}
	\caption{(a) The projected band structure of ZrCl$_2$. (b) Projected density of states of the Zr $d$ orbitals following trigonal prismatic crystal-field splitting in monolayer ZrCl$_2$.}
	\label{fig.s6}
\end{figure}

\begin{figure}[htb]
	\centering
	\includegraphics[width=0.8\columnwidth]{./pics7.pdf}
	\caption{3D view of the conduction-band energy dispersion near the CBM of monolayer ZrCl$_2$. The inset shows the Mexican-hat-like feature close to the CBM, suggesting exit of Van Hove singularities.}
	\label{fig.s7}
\end{figure}

\clearpage

\subsection*{3. Superconductivity in electron-doped ZrX$_2$}

\begin{figure}[htbp]
	\centering
	\includegraphics[width=0.8\columnwidth]{./pics8.pdf}
	\caption{Band structure and density of states of monolayer (a) ZrBr$_2$ and (c) ZrI$_2$ at $x = 0.01$ \textit{e}/f.u.. The corresponding phonon dispersion, PHDOS, Eliashberg spectral function $\alpha^2F(\omega)$, and EPC constant $\lambda$ of monolayer (c) ZrBr$_2$ and (d) ZrI$_2$ at at the same doping level. }
	\label{fig.s8}
\end{figure}

\begin{figure}[htbp]
	\centering
	\includegraphics[width=0.5\columnwidth]{./pics9.pdf}
	\caption{Superconducting $T_c$ of monolayer ZrX$_2$ (X = Cl, Br, I), as a function of electron doping concentration, calculated using both the McMillan-Allen-Dynes (MAD) and Allen-Dynes modified (AD) formulas.}
	\label{fig.s9}
\end{figure}

\subsection*{4. Superconductivity in hole-doped ZrBr$_2$ and ZrI$_2$}

\begin{figure}[htbp]
	\centering
	\includegraphics[width=0.9\columnwidth]{./pics10.pdf}
	\caption{Phonon dispersion, PHDOS, Eliashberg spectral function $\alpha^2F(\omega)$, and EPC constant $\lambda$ of monolayer (a) ZrBr$_2$ and (b) ZrI$_2$ at $x = 0.25$ h/f.u..}
	\label{fig.s10}
\end{figure}

\begin{figure}[htbp]
	\centering
	\includegraphics[width=0.8\columnwidth]{./pics11.pdf}
	\caption{Band structure, density of states and Fermi surface of monolayer ZrBr$_2$ at (a) $x = 0.2$ \textit{h}/f.u. and (b) $x = 0.25$ \textit{h}/f.u. Corresponding results for monolayer ZrI$_2$ at $x = 0.2$ \textit{h}/f.u. and (d) $x = 0.25$ \textit{h}/f.u.}
	\label{fig.s11}
\end{figure}

\clearpage

\subsection*{5. Interstitial electrons  states and superconductivity}

\begin{figure}[htb]
	\centering
	\includegraphics[width=0.9\columnwidth]{./pics12.pdf}
	\caption{(a) 3D ELF with iso-surface set to 0.7 a.u., (b) 2D ELF heatmap in the ab-plane at c = 0.5, and (c) the 1D ELF along the path from (1, 0, 0.5) to (0, 1, 0.5) (black line in (b)). Corresponding results for monolayer ZrBr$_2$ and ZrI$_2$ are shown in (d)(e)(f) and (g)(h)(f), respectively.}
	\label{fig.s12}
\end{figure}

\begin{table*}[htbp]
	\centering
	\caption{Interstitial non-nuclear attractor (NNA) in monolayer ZrCl$_2$ obtained from BadELF and Critic2 analyses. The N$_{\text{NNA}}$ is the integrated electron count within the ELF zero-flux basin. $\nabla^{2} \rho (\textbf{r}_{0})$ is evaluated at the Critic2 NNA center. Both methods identify a single NNA satisfying the electride criteria \cite{electride1} .}
	\begin{tabular*}{\textwidth}{@{\extracolsep{\fill}}ccccc}
		\hline
		Methods & NNA site (fractional) & N$_{\text{NNA}}$ (\textit{e}/f.u.) & NNA ELF$_{\text{MAX}}$ (a.u.) & $\nabla^{2} \rho (\textbf{r}_{0})$ (\textit{e}/Bohr$^{5}$) \\
		\hline
		BadELF \cite{badelf} & (0.32, 0.66, 0.50) & 0.959 & 0.874 & --- \\
		Critic2 \cite{cc1,cc2} & (0.34, 0.66, 0.50) & 0.918 & 0.874 & -0.00768 \\
		\hline
	\end{tabular*}
	\label{tab.s1}
\end{table*}

\begin{figure}[htbp]
	\centering
	\includegraphics[width=0.8\columnwidth]{./pics13.pdf}
	\caption{Orbital-resolved band structures of monolayer ZrCl$_2$ projected onto (a) Zr $d_{xy}/d_{x^2-y^2}$, (b) Zr $d_{z^2}$, and (c) the interstitial quasi-atom Q, respectively. (d) Total and projected density of states of monolayer ZrCl$_2$. Here, the quasi-atom Q locates at the interstitial site (0.33, 0.67, 0.5).}
	\label{fig.s13}
\end{figure}

\begin{figure}[htbp]
	\centering
	\includegraphics[width=0.8\columnwidth]{./pics14.pdf}
	\caption{Ratio of the PDOS at the Fermi level from the (a) interstitial Q and (b) Zr-$d_{z^2}$ to the total DOS as a function of hole doping concentration for monolayer ZrCl$_2$.}
	\label{fig.s14}
\end{figure}

\clearpage

\subsection*{6. Magnetism in hole-doped ZrX$_2$}

\begin{figure}[htbp]
	\centering
	\includegraphics[width=0.9\columnwidth]{./pics15.pdf}
	\caption{(a) Ferromagnetic (FM), (b) stripe antiferromagnetic (sAFM), and (c) zigzag antiferromagnetic (zAFM) configurations for monolayer ZrX$_2$. Red and blue arrows denote spin-up and spin-down, respectively. (d) Total energy difference $\Delta$E = E$_{\text{FM}}$ − E$_{\text{sAFM}}$ (meV/f.u.) as a function of hole doping concentration \textit{x} (\textit{h}/f.u.) for ZrCl$_2$, ZrBr$_2$, and ZrI$_2$. Here, the positive $\Delta$E indicates that the sAFM state is more stable, while negative $\Delta$E indicates that the FM state is more stable. The zAFM initialization converges to a nonmagnetic state, indicating that this configuration is not stable.}
	\label{fig.s15}
\end{figure}

\clearpage

\subsection*{7. Phonon stability and phase diagram under full relaxation}

\begin{figure}[htbp]
	\centering
	\includegraphics[width=0.75\columnwidth]{./pics16.pdf}
	\caption{Phonon dispersions of ZrCl$_2$ calculated for structures with fixed lattice constants at (a) 0.3 \textit{e}/f.u. and (b) 0.35 \textit{e}/f.u., and for fully relaxed structures at (c) 0.3 \textit{e}/f.u. and (d) 0.35 \textit{e}/f.u., respectively.}
	\label{fig.s16}
\end{figure}

\begin{figure}[htbp]
	\centering
	\includegraphics[width=0.8\columnwidth]{./pics17.pdf}
	\caption{The full relaxed lattice constants \textit{a} (Å) and phase transition of ZrCl$_2$ as a function of doping concentration \textit{x}. Here, the positive/negative value reflect hole/electron doping.}
	\label{fig.s17}
\end{figure}

\bibliography{./SupplementRef}